\documentclass[12pt]{article}
\usepackage[utf8]{inputenc} 
\usepackage[T1]{fontenc}    
\usepackage{hyperref}       
\usepackage{url}            
\usepackage{booktabs}       
\usepackage{amsfonts}       
\usepackage{nicefrac}       
\usepackage{microtype}      
\usepackage{amsmath}
\usepackage{graphicx}
\usepackage{subfig}
\usepackage{enumitem}
\usepackage{indentfirst}
\usepackage{multirow}
\usepackage{libertine}
\usepackage{gensymb}
\usepackage[english]{babel}
\usepackage[top=2.5cm, bottom=2.5cm, left=2.5cm, right=2.5cm]{geometry}
\usepackage[svgnames]{xcolor}
\usepackage{xy}
\xyoption{all}
\usepackage{natbib}

\newtheorem{definition}{Definition}

\usepackage{color}

\title{Conditional Inferences Based on Vine Copulas with Applications to Credit Spread Data of Corporate Bonds}
\author{
	Shenyi Pan\thanks{Department of Statistics, University of British Columbia, Vancouver, BC Canada V6T 1Z4. Email: \href{mailto:shenyi.pan@stat.ubc.ca}{shenyi.pan@stat.ubc.ca}.}
	\and
	Harry Joe\thanks{Department of Statistics, University of British Columbia, Vancouver, BC Canada V6T 1Z4. Email: \href{mailto:harry,joe@ubc.ca}{Harry.Joe@ubc.ca}.}
	\and 
	Guofu Li\thanks{AI Lab, Ping An Asset Management Co., Ltd, No. 1333 Lujiazui Huan Road, Shanghai, China. Email: \href{mailto:li.guofu.l@gmail.com}{li.guofu.l@gmail.com}.}
}
\date{}

\begin{document}

\maketitle

\begin{abstract}
Understanding the dependence relationship of credit spreads of corporate
bonds is important for risk management.  Vine copula models with tail
dependence are used to analyze a credit spread dataset of Chinese
corporate bonds, understand the dependence among different sectors and
perform conditional inferences.  It is shown how the effect of tail
dependence affects risk transfer, or the conditional distributions
given one variable is extreme.  Vine copula models also provide more
accurate cross prediction results compared with linear regressions. These
conditional inference techniques are a statistical contribution for
analysis of bond credit spreads of investment portfolios consisting of
corporate bonds from various sectors.
\end{abstract}


\section{Introduction} \label{sec: intro}

For multivariate financial data from different companies or markets,
such as stock log-returns, changes in daily bond yields and CDS spreads,
tail dependence (possibly asymmetric) can be seen from scatter plots and 
this property is taken into account in a dependence model.
Some relevant references are
\cite{okimoto2008new}, \cite{zhang2014vine}, \cite{weiss2015mixture},
\cite{oh.patton2018time-varying}, \cite{duan2019model}, \cite{nguyen2019parallel}, \cite{kim2020modeling} and
\cite{krupskii2020flexible}. 
If a dependence model using a vine or factor copula is fitted to this type
of multivariate financial data, conditional inferences such as
stress testing or risk transfer study (effects on other variables conditioned on one variable being extreme) and prediction of one variable given others can be made.
The effects of inferences from a dependence model with tail dependence can be 
seen from comparisons with a Gaussian dependence model;
examples are in \cite{brechmann2013conditional} and
\cite{krupskii2020flexible}.

We study a type of financial data which consist of average 
credit spreads of corporate bonds from different sectors in China.
The series of daily credit spread changes are GARCH-filtered. Diagnostics
suggest that these transformed variable have pairwise tail dependence for
most pairs of sectors.
A vine copula is fitted to the GARCH-filtered series, and the (conditional)
dependence can be interpreted to show which sectors are more strongly
related. Some conditional inferences such as risk transfer from one
sector to others show the effect of tail dependence in the dependence model.
For credit spread data across different sectors, it is often of interest to model the dependence structure among sectors, such that sectors of central importance can be identified and how upside or downside changes in credit spreads propagate through sectors can be analyzed.
Moreover, corporate bonds are one of the primary types of financial assets purchased by Chinese asset management companies, and credit spreads are the most important metric in bond pricing and risk management. Widening credit spreads indicate growing concern on the ability of corporate borrowers to repay bondholders on the maturity date, while narrowing credit spreads indicate improving corporate creditworthiness. Therefore, accurate prediction of sector-wise credit spreads plays a vital role in guiding corporate bond investment decisions. It also contributes a statistical way to actively monitor the quality of the bond portfolio in addition to analyzing the public bond ratings, whose adjustments often lag behind the events that affect the credit quality of corporate bonds.

For multivariate observational data, different conditional distributions
can be considered after fitting a multivariate dependence model.
In the stock or bond market, an investment portfolio usually consists of a large number of stocks or bonds. The daily returns of all the stocks or bonds are simultaneously observed and equally important in affecting the return performance of a portfolio. In these cases, it is of more interest to model the underlying dependence structure based on the concurrently observed data. More formally, assume a set of variables $\boldsymbol{X} = (X_1, \dots, X_d)$ are measured simultaneously. A natural approach is to fit a joint distribution to $(X_1, \dots, X_d)$ given an observed sample $(x_{i1}, \dots, x_{id})$ for $i = 1, \dots, n$; this joint distribution can be used to infer the dependence structure among variables. 

For an index $j\in\{1,\ldots,d\}$,
let $\boldsymbol{X}_{-j}$ represents all variables excluding $X_j$.
For conditional inference regarding cross predictions and risk transfer,
one could consider the conditional distribution of $X_j$ given $\boldsymbol{X}_{-j}$ or the conditional distribution of $\boldsymbol{X}_{-j}$ given that
$X_j$ is extreme.
For the former,
the conditional expectation $\mathbb{E}(X_j |\boldsymbol{X}_{-j})$ and conditional quantiles $F^{-1}_{X_j |\boldsymbol{X}_{-j}}(p|\boldsymbol{x}_{-j})$ can be obtained from the conditional distribution for performing cross predictions. This becomes the usual multiple regression if the joint distribution of $\boldsymbol{X}$ is multivariate Gaussian. Nevertheless, multiple regression directly models the conditional distributions $F_{X_j |\boldsymbol{X}_{-j}}(x_j |\boldsymbol{x}_{-j})$. As a result, $d$ models for the conditional distributions need to be separately fitted to perform cross predictions for $d$ variables. This may cause problems for large $d$ in practice. In contrast, if one models the joint distribution of $\boldsymbol{X}$, the $d$ conditional distributions can be derived from the joint distribution to perform cross predictions. As a result, only one model is required to be maintained and updated. Unlike multiple regression, prediction based on the joint distribution uses information on the distributions of the variables and does not specify a simple linear or polynomial equation for the conditional expectation.

For conditional inference to work well, the joint distribution must be
approximated well. In recent years,
the vine pair-copula construction has proven to be a flexible tool in high-dimensional non-Gaussian dependence modeling. In a vine copula model, vine graphs represented by a sequence of connected trees are adopted to specify the dependence structure, while bivariate copulas are used as the basic building blocks on the edges of vines; see for example,
\cite{bedford2002vines}, \cite{aas2009pair}, and \cite{kurowicka2011dependence}.

The possibility of applying copula models to make predictions for a predetermined response variable has been explored in previous literature. For example, \cite{parsa2011copula} use a multivariate Gaussian copula to model the joint distribution and derive a closed-form conditional distribution for prediction. \cite{noh2015semiparametric} perform semiparametric conditional quantile regression through copula-based multivariate models. Vine copulas are used by \cite{kraus2017d} and \cite{schallhorn2017d} to perform quantile regression for continuous data and mixed discrete-continuous data, respectively, but the vine structure is restricted to a boundary class of vines called the drawable vines (D-vines). 
\cite{chang2019prediction} propose a unified algorithm to handle mixed discrete-continuous data with regular vines (R-vines). Nevertheless, all these existing papers assume that there is a predetermined response variable, and the conditional distribution of the response variable given the other variables is of primary interest. 

In this paper, we include an algorithm to compute $d$ conditional distributions 
of one variable given the other variables for cross prediction
from a single joint distribution.
Compared with commonly used prediction methods such as linear regression or random forest, cross prediction based on R-vines only needs to maintain one model for the joint distribution instead of $d$ separate models; this is more feasible and efficient in practice. Compared with multiple linear regression, R-vine copulas are more flexible. Various shapes of conditional quantiles of one variable given the others can be obtained depending on how pair-copulas are chosen on the edges of the vine. For the credit spread data of Chinese corporate bonds across different sectors,
the cross prediction method is applied and compared with Gaussian-based
methods. Risk transfer is also considered when one centrally
dependent sector is set to have extreme GARCH-filtered values.
This shows the effect of tail dependence on sectors that are more closely
related to the centrally chosen sector.

The remaining sections are organized as follows. Section~\ref{sec: vine_cop} gives an overview of copula models and vine copulas. 
New material on conditional inference from vine copulas is presented
in Section~\ref{sec: conditional}; this includes the algorithms to
perform cross prediction and conditional simulation for the risk transfer study.
Comparison criteria to assess different models for cross prediction
are summarized in Section~\ref{sec: sim} and some models are compared
in a brief simulation study.
Section~\ref{sec: data} gives a brief introduction to the credit spread dataset of Chinese corporate bonds as well as the required processing steps. 
Conditional inferences are considered in Section~\ref{sec: app}.
Section~\ref{sec: conc} consists of concluding discussions.

\section{Vine Copula Models} \label{sec: vine_cop}

In this section, an overview of copula models is provided. Section~\ref{subsec: copula} has some basic results for bivariate and multivariate copulas. Section~\ref{subsec: vine} gives a brief introduction to the vine structure for conditional bivariate dependence.
Section~\ref{subsec: vine_struct} summarizes some methods for deciding
on a vine structure and
Section~\ref{subsec: biv_cop_select} discusses an approach to selecting
parametric bivariate copula families on the edges of a vine.

For fitting a vine copula, we assume that
the dataset consists of $(x_{i1}, \dots, x_{id})$ for $i=1, \dots, n$; these vectors are considered as independent realizations of a continuous random vector $(X_1, \dots, X_d)$.
The procedure in the latter two subsections assumes that the $d$ variables are monotonically
related to each other, so that dependence can be summarized by a
matrix of Spearman or van der Waerden correlations (after variables have
been individually transformed to follow $U(0,1)$ or $N(0,1)$ distribution, respectively).
After fitting univariate distributions to individuals variables,
probability integral transforms are applied to convert them to transformed
observation vectors in $[0,1]^d$.

\subsection{Introduction to Copula} \label{subsec: copula}

A copula is a multivariate distribution function with univariate Uniform(0,1) margins. Sklar's theorem (\cite{sklar1959fonctions}) decomposes a $d$-dimensional distribution into two parts: the marginal distributions and the copula function linking the margins. It states that for a $d$-variate distribution $F\in\mathcal{F}(F_1,\dots,F_d)$, with $j$th univariate margin $F_j$, the copula associated with $F$ is a distribution function $C: [0,1]^d\rightarrow[0,1]$ with $U(0,1)$ margin that satisfies 
\[
F(\boldsymbol{x})=C(F_1(x_1),\dots,F_d(x_d)),\ \boldsymbol{x}\in\mathbb{R}^d;
\]
Conversely, if $F$ is a continuous $d$-variate distribution function with univariate margins $F_1, \dots, F_d$, and quantile functions $F_1^{-1},\dots,F_d^{-1}$, then the copula
\[
C(\boldsymbol{u})=F(F_1^{-1}(u_1),\dots,F^{-1}(u_d)),\ \boldsymbol{u}\in[0,1]^d,
\]
is the unique choice. Detailed introductions to multivariate copula constructions can be found in \cite{joe2014dependence} 
and \cite{czado2019analyzing}.

Vine copulas or pair-copula constructions are constructed from a sequence of bivariate copulas.
Some results for bivariate copulas that are used in subsequent sections are summarized here. Consider two continuous random variables $X_1$ and $X_2$ with joint cumulative density function (CDF) $F_{12}(x_1, x_2)$, marginal CDFs $F_1$, $F_2$, and probability density function (PDF) $f_{12}(x_1, x_2)$. According to Sklar's theorem, there exists a copula $C(u_1, u_2)$ such that $F_{12}(x_1, x_2) = C(F_1(x_1), F_2(x_2))$.

The copula density of $C(u_1, u_2)$ is
$ c(u_1, u_2) = \partial^2 C(u_1, u_2)/\partial u_1 \partial u_2$.
The conditional CDF of $U_1$ given $U_2 = u_2$ is
\[
C_{1|2}(u_1|u_2) = \mathbb{P}(U_1 \leq u_1|U_2 = u_2) = \frac{\partial C(u_1, u_2)}{\partial u_2}.
\]
The conditional quantile function $C_{1|2}^{-1}(\cdot|u_2)$ is the inverse function of $C_{1|2}(\cdot|u_2)$. Exchanging $U_1$ and $U_2$, the conditional CDF $C_{2|1}(\cdot|u_1)$ and quantile function $C_{2|1}^{-1}(\cdot|u_1)$ can be defined similarly.

Let $f_1$, $f_2$, and $f_{12}$ be the density functions of $X_1$, $X_2$ and $(X_1, X_2)$, respectively. The joint density function $f_{12}$ can be decomposed as
\[
f_{12}(x_1, x_2) = c(F_1(x_1), F_2(x_2))f_1(x_1)f_2(x_2).
\]

\subsection{Overview of Vine Structure} \label{subsec: vine}

A regular vine (R-vine) is a nested set of trees where the edges in the first tree are the nodes of the second tree, the edges of the second tree are the nodes of the third tree, etc. Vines are useful in specifying the dependence structure for general multivariate distributions on $d$ variables with edges in the first tree representing
pairwise dependence and edges in subsequent trees representing conditional dependence.

The first tree in a vine represents $d$ variables as nodes and the bivariate dependence of $d-1$ pairs of variables as edges. The second tree describes the conditional dependence of $d-2$ pairs of variables conditioned on another variable; nodes are the edges in tree 1, and a pair of nodes can be connected if
there is a common variable in the pair. The third tree describes the conditional dependence of $d-3$ pairs of variables conditioned on two other variables; nodes are the edges in tree 2, and a pair of nodes could be connected if there are two common conditioning variables in the pair. This continues until tree $d-1$ has only one edge that describes the conditional dependence of two variables conditioned on the
remaining $d-2$ variables.

\begin{figure}[!ht] 
	\centering
	\xymatrix{
		& *+[F]{1} \ar@{-}[rr]^{12} &&
		*+[F]{2} \ar@{-}[rr]^{23} &&  *+[F]{3} \ar@{-}[rr]^{34} && *+[F]{4} \ar@{-}[rr]^{45} && *+[F]{5} & T_1 \\ 
		&& *+[F]{12} \ar@{-}[rr]^{13;2} && *+[F]{23} \ar@{-}[rr]^{24;3} && *+[F]{34} \ar@{-}[rr]^{35;4} && *+[F]{45} & T_2 \\
		&&& *+[F]{13;2} \ar@{-}[rr]^{14;23} && *+[F]{24;3} \ar@{-}[rr]^{25;34} && *+[F]{35;4} & T_3 \\
		&&&& *+[F]{14;23} \ar@{-}[rr]^{15;234} && *+[F]{25;34} & T_4 \\
	}
	\caption{A D-Vine structure on five variables.}
	\label{fig: vine_example}
\end{figure}
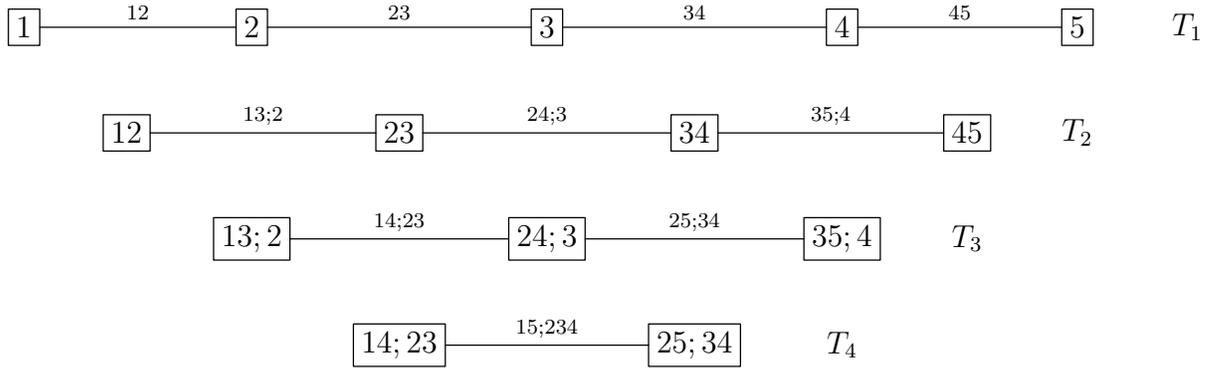

The mathematical definition of a vine as a sequence of trees is given in \cite{bedford2002vines}. The formal definition is given as follows:
\begin{definition}
	\label{def: vine}
	$V$ is a vine on $d$ variables, with $E(V) = \bigcup_{i = 1}^{d-1} E(T_\ell)$ denoting the set of edges of $V$ if
	\setlist{nolistsep}
	\begin{enumerate}[noitemsep]
		\item
		$V = (T_1, \ldots, T_{d-1})$; 
		
		\item
		$T_1$ is a tree with nodes $N(T_1) = \{1, 2, \ldots, d\}$, and edges $E(T_1)$. For $\ell > 1$,
		$T_{\ell}$
		is a tree with nodes $N(T_{\ell}) = E(T_{\ell - 1})$;
		
		\item
		(proximity condition) For $\ell = 2, \ldots, d -1$, for $\{n_1, n_2\} \in E(T_{\ell})$, $\# (n_1 \triangle n_2) = 2$, where $\triangle$ denotes symmetric difference and $\#$ denotes cardinality.
		
	\end{enumerate}
\end{definition}
An R-vine can be represented by the edge sets at each level $E(T_\ell)$ or by a graph.
A vine array is a compact method to encode the conditional distributions used in a vine. A vine array $A = (a_{\ell j})_{\ell = 1, \dots, d; j = \ell, \dots, d}$, for an R-vine $V = (T_1, \dots, T_{d-1})$ on $d$ elements is a $d\times d$ upper triangular matrix. It satisfies the following two conditions:
\begin{itemize}
	\item The diagonal of $A$ is a permutation of $(1, \dots, d)$.
	\item For $j = 2, \dots, d$, the $j$th column has $(a_{1j}, \dots, a_{j-1,j})$ being a permutation of $(a_{11}, \dots, a_{j-1,j-1})$. In the first row, $a_{1j}$ can be any element in $\{a_{11}, \dots , a_{j-1,j-1}\}$. For $\ell = 2, \dots, j-1$, the set $\{a_{1j}, \dots, a_{\ell j}\}$ is equal to $\{a_{1k}, \dots, a_{\ell - 1, k}, a_{kk}\}$ for at least one $k$ in columns $\ell, \dots, j - 1$.
\end{itemize}
For $\ell=2,\ldots,d-1$ and $\ell<j\le d$,
row $\ell$ and column $j$ of the vine array indicates that the variable $a_{\ell j}$
is connected to the variable $a_{jj}$ in tree $T_\ell$, conditioned on variables $a_{1j}, \dots, a_{\ell -1,j}$. In other words, the first $\ell$ rows of $A$ and the diagonal elements encode the $\ell$th tree $T_\ell$, such that
the edge  $[a_{\ell j}, a_{jj}|a_{1j}, \dots , a_{\ell-1,j}] \in E(T_\ell)$ summarizes the conditional dependence on $\ell-1$ variables, for $\ell + 1 \leq j \leq d$. As an illustrative example, the vine array $A_1$
\[
A_1 = 
\begin{bmatrix}
1 & 1 & 2 & 3 & 4\\
& 2 & 1 & 2 & 3\\
&   & 3 & 1 & 2\\
&   &   & 4 & 1\\
&   &   &   & 5\\
\end{bmatrix}
\]
represents the D-vine structure (a boundary class of vines) in Figure~\ref{fig: vine_example}. In each tree of a D-vine, there is a path passing through all the nodes in that tree. Note that \cite{dissmann2013selecting} use the vine array in reverse row and column indexing.

To get a multivariate distribution from a vine, for each edge in the vine, there is a bivariate copula assigned to it. Specifically, bivariate distributions are assigned to edges on the first tree and bivariate conditional distributions are assigned to edges on the subsequent trees. For a vine array, there should be a bivariate copula associated with each of the $d(d - 1)/2$ entries in the upper diagonal. The bivariate copula corresponding to $a_{\ell j}$ for $1 \leq l < j \leq d$ is denoted by $C_{a_{\ell j}, a_{jj}; S_{\ell j}}$ with density function $c_{a_{\ell j}, a_{jj}; S_{\ell j}}$, where $S_{\ell j} = \{a_{1j},\dots, a_{\ell-1,j}\}$ is the conditioning set for this position. Note that $S_{\ell j} = \emptyset$ if $\ell = 1$ and in tree 1,
$C_{a_{1j}, a_{jj}}$ summarizes the dependence of a pair of variables for $j=2,\ldots,d$. 
In tree $\ell\in \{2,\ldots,d-1\}$, the bivariate copula $C_{a_{\ell j}, a_{jj}; S_{\ell j}}$ is used to link the conditional distributions $F_{a_{\ell j}|S_{\ell j}}$ and $F_{a_{jj}|S_{\ell j}}$, 
and it summarizes the conditional dependence of variables indexed
as $a_{\ell j}$ and $a_{jj}$ given the variables in the index set $S_{\ell j}$.

Let $f_1, \dots, f_d$ be the univariate densities. The joint density of $(X_1, \dots , X_d)$ based on the vine structure specified by a vine array can be decomposed as
\begin{equation}
f_{1:d}(x_1, \dots, x_d) = \prod_{i=1}^{d}f_i(x_i) \cdot \prod_{\ell = 1}^{d-1} \prod_{j = \ell + 1}^{d}c_{a_{\ell j}, a_{jj}; S_{\ell j}}\left[F_{a_{\ell j}|S_{\ell j}}\left(x_{a_{\ell j}}|\boldsymbol{x}_{S_{\ell j}}\right), F_{a_{jj}|S_{\ell j}}\left(x_{a_{jj}}|\boldsymbol{x}_{S_{\ell j}}\right)\right], \label{eq: vine_density}
\end{equation}
where the conditional distributions $F_{a_{\ell j}|S_{\ell j}}$ and $F_{a_{jj}|S_{\ell j}}$ are determined in a recursive manner, using bivariate copulas on the edges in previous trees.
Derivations of this result are given in \cite{bedford2001probability},
\cite{joe2014dependence} and \cite{czado2019analyzing}.

Since a joint distribution can be decomposed into univariate marginal distributions and a dependence structure among variables, estimation can proceed in a two-stage manner. The first step estimates the univariate marginal distributions $\widehat{F}_j$ for $j = 1, \dots, d$. The u-scores vectors $(u_{i1},\ldots,u_{id})$, $i=1, \dots, n$, are obtained by applying the probability integral transform: 
\begin{equation}
 u_{i,j} = \widehat{F}_j(x_{i,j}). \label{eq: PIT} 
\end{equation}
The second step fits a vine copula model based on the u-score vectors. 
In our approach, the second step involves two components: finding a vine structure describing the underlying dependence and deciding which bivariate
parametric copula families to use on the edges of the vine. 

\subsection{Vine Structure Learning} \label{subsec: vine_struct}

Learning the optimal structure of a vine is computationally intractable in general. There are a large number of possible vine structures which result in a huge search space for a high-dimensional dataset if one would like to find the optimal one. 

\cite{dissmann2013selecting} propose a greedy method based on the maximum spanning tree (MST) algorithms with different choices for edge weights that reflect the strength of the dependence between pairs of variables. 
Parametric bivariate copulas are fitted to tree $\ell$ before deciding on a
tree to represent conditional dependence for tree $\ell+1$.
The trees of the vine are sequentially constructed by maximizing the sum of the edge weights at each tree level.

Alternatively, assuming variables are monotonically related,
an entire vine structure can be decided based on heuristics
of having pairs of variables with stronger dependence and conditional
dependence in lower trees; this is
followed by deciding on the bivariate copulas on the edges of the vine.
\cite{brechmann2015truncation}, using a genetic algorithm and neighbors of
maximum spanning trees, explore the space of truncated vines based on conditional dependence from the van der Waerden correlation matrix. 
More recently, \cite{chang2019vine} propose to learn vine structures using the Monte Carlo tree search algorithm to explore a larger search space of possible vines based on the van der Waerden correlation matrix.
Various algorithms have been proposed based on different heuristics but no method is expected to perform universally the best.

\subsection{Bivariate Copula Selection} \label{subsec: biv_cop_select}

After fitting the univariate marginal distributions and finding the vine structure, 
parametric bivariate copulas can be fitted sequentially on each edge from $T_1$ to $T_{d-1}$. Suppose there are $M$ bivariate copula candidate families for each edge in the vine.

In the first tree, consider the edge $[a_{1j},a_{jj}]$. The log-likelihood of the bivariate copula model $(m)$ on this edge is
\[
\mathcal{L}_{a_{1j}, a_{jj}}\left(\boldsymbol{\theta}^{(m)}\right) = \sum_{i = 1}^{n} \log c^{(m)}_{a_{1j}, a_{jj}}\bigl[u_{i, a_{1j}}, u_{i, a_{jj}}; \boldsymbol{\theta}^{(m)} \bigr].
\]
Commonly used model selection criteria include Akaike information criterion (AIC) and Bayesian information criterion (BIC). For the edge $[a_{1j},a_{jj}]$, there are $M$ AIC and BIC values. The corresponding AIC and BIC values for model $(m)$ are defined as:
\begin{align*}
&\text{AIC}^{(m)}_{a_{1j}, a_{jj}}\left(\boldsymbol{\theta}^{(m)}\right) = -2\mathcal{L}_{a_{1j}, a_{jj}}\left(\boldsymbol{\theta}^{(m)}\right) + 2\text{card}\left(\boldsymbol{\theta}^{(m)}\right) \mbox{ and}\\
&\text{BIC}^{(m)}_{a_{1j}, a_{jj}}\left(\boldsymbol{\theta}^{(m)}\right) = -2\mathcal{L}_{a_{1j}, a_{jj}}\left(\boldsymbol{\theta}^{(m)}\right)+ \log (n)\text{card}\left(\boldsymbol{\theta}^{(m)}\right),
\end{align*}
where $\text{card}(\boldsymbol{\theta}^{(m)})$ represents the cardinality of the copula parameter vector.  For each candidate parametric bivariate copula model on an edge, the maximum likelihood parameter estimator $\widehat{\boldsymbol{\theta}}^{(m)}$ is obtained. The parametric copula model with the lowest AIC or BIC is selected for that edge. 

In tree $\ell\in \{2,\ldots,d-1\}$, consider the edge $[a_{\ell j},a_{jj};S_{\ell j}]$. The log-likelihood of the bivariate copula model $(m)$ on this edge is
\[
\mathcal{L}_{a_{\ell j}, a_{jj}; S_{\ell j}}\left(\boldsymbol{\theta}^{(m)}\right) = \sum_{i = 1}^{n} \log c^{(m)}_{a_{\ell j}, a_{jj}; S_{\ell j}}\left[C_{a_{\ell j}|S_{\ell j}}\left(u_{i, a_{\ell j}}|\boldsymbol{u}_{i, S_{\ell j}}\right), C_{a_{jj}|S_{\ell j}}\left(u_{i, a_{jj}}|\boldsymbol{u}_{i, S_{\ell j}}\right); \boldsymbol{\theta}^{(m)} \right],
\]
where $\boldsymbol{u}_{i, S_{\ell j}} = \{u_{ik}: k\in S_{\ell j}\}$. The pseudo observations $C_{a_{\ell j}|S_{\ell j}}\left(u_{i, a_{\ell j}}|\boldsymbol{u}_{i, S_{\ell j}}\right)$ and $C_{a_{jj}|S_{\ell j}}\left(u_{i, a_{jj}}|\boldsymbol{u}_{i, S_{\ell j}}\right)$ are obtained based on the fitted copulas at the previous levels. The corresponding AIC and BIC values for model $(m)$ are defined as:
\begin{align*}
	&\text{AIC}^{(m)}_{a_{\ell j}, a_{jj}; S_{\ell j}}\left(\boldsymbol{\theta}^{(m)}\right) = -2\mathcal{L}_{a_{\ell j}, a_{jj}; S_{\ell j}}\left(\boldsymbol{\theta}^{(m)}\right) + 2\text{card}\left(\boldsymbol{\theta}^{(m)}\right) \mbox{ and}\\
	&\text{BIC}^{(m)}_{a_{\ell j}, a_{jj}; S_{\ell j}}\left(\boldsymbol{\theta}^{(m)}\right) = -2\mathcal{L}_{a_{\ell j}, a_{jj}; S_{\ell j}}\left(\boldsymbol{\theta}^{(m)}\right)+ \log (n)\text{card}\left(\boldsymbol{\theta}^{(m)}\right).
\end{align*} 

This approach to selecting bivariate copulas is implemented in the \verb|VineCopula| R package (\cite{schepsmeier2019vinecopula}).

\section{Conditional Inference} \label{sec: conditional}

This section discusses the theory and algorithms for conditional inference
with the financial data example introduced in Section \ref{sec: data}. 
The details for implementing this are not in the references for vine copulas
given in Section~\ref{sec: vine_cop}.

Section~\ref{subsec: cross_pred} describes an algorithm that derives $d$ conditional distributions from the joint density function of the vine copula and allows cross predictions for each variable given the others.
Simulation from a vine copula conditioned on a variable with a given
value is covered in Section~\ref{subsec: conditionalsimulation}.
This is useful in order to quantify how extreme values of one variable affect
other variables.

\subsection{Cross Prediction} \label{subsec: cross_pred}

The inputs of cross prediction are a vine copula model with a vine array $A = (a_{\ell j})$, a vector of new observations $\widetilde{\boldsymbol{x}} = (\widetilde{x}_1, \dots, \widetilde{x}_d)$, as well as a percentile $q\in (0,1)$. The outputs of cross prediction are the conditional quantiles $F_{j|\{(1:d)\setminus j\}}^{-1}(q|\widetilde{\boldsymbol{x}}_{\{(1:d)\setminus j\}})$ for $j = 1, \dots, d$, where $\{(1:d)\setminus j\}$ denotes the sequence from 1 to $d$ excluding the element $j$.

Based on Equation~\eqref{eq: vine_density}, the joint density of $(X_1, \dots, X_d)$ given the vine array $A$ is
\begin{align*}
f_{1:d}(x_1, \dots, x_d) &= \left\{\prod_{i=1}^{d}f_i(x_i)\right\} \left\{\prod_{\ell = 1}^{d-1} \prod_{j = \ell + 1}^{d}c_{a_{\ell j}, a_{jj}; S_{\ell j}}\left[F_{a_{\ell j}|S_{\ell j}}\left(x_{a_{\ell j}}|\boldsymbol{x}_{S_{\ell j}}\right), F_{a_{jj}|S_{\ell j}}\left(x_{a_{jj}}|\boldsymbol{x}_{S_{\ell j}}\right)\right]\right\} \\
&= \left\{\prod_{i=1}^{d}f_i(x_i)\right\} \left\{\prod_{\ell = 1}^{d-1} \prod_{j = \ell + 1}^{d}c_{a_{\ell j}, a_{jj}; S_{\ell j}}\left[C_{a_{\ell j}|S_{\ell j}}\left(u_{a_{\ell j}}|\boldsymbol{u}_{S_{\ell j}}\right), C_{a_{jj}|S_{\ell j}}\left(u_{a_{jj}}|\boldsymbol{u}_{S_{\ell j}}\right)\right]\right\} \\
&\equiv \left\{\prod_{i=1}^{d}f_i(x_i)\right\}c_{1:d}(u_1, \dots, u_d),
\end{align*}
where $u_{j}=F_j(x_j)$ for $j=1,\ldots,d$.
The first term is the product of the univariate densities, which can be evaluated using observations of $\boldsymbol{X}$ with the univariate marginal distributions. The second term is the copula density, which can be evaluated using u-scores $\boldsymbol{u}_i$ (as given in Equation~\eqref{eq: PIT}) with the fitted vine copula model.

Deriving the conditional distribution of the two variables $X_{a_{d-1,d}}$ and $X_{a_{dd}}$ conditioned on $\boldsymbol{X}_{S_{d-1,d}}$ at the last level does not involve any integration. Based on the bivariate copula model on the edge in $T_{d-1}$, one can get
\begin{eqnarray*}
\lefteqn{ F_{a_{d-1,d}|S_{d-1,d}\bigcup a_{dd}}\left(x_{a_{d-1,d}}|\boldsymbol{x}_{S_{d-1,d}}, x_{a_{dd}}\right)} \\
&=& C_{a_{d-1,d}|S_{d-1,d}\bigcup a_{dd}}\left(u_{a_{d-1,d}}|\boldsymbol{u}_{S_{d-1,d}}, u_{a_{dd}}\right) \\
&=& C_{a_{d-1,d}|a_{dd};S_{d-1,d}}\left(C_{a_{d-1,d}|S_{d-1,d}}(u_{a_{d-1,d}}|\boldsymbol{u}_{S_{d-1,d}})|C_{a_{dd}|S_{d-1,d}}(u_{a_{dd}}|\boldsymbol{u}_{S_{d-1,d}})\right),
\end{eqnarray*}
where $C_{a_{d-1,d}|a_{dd};S_{d-1,d}}(v|w) = {\partial C_{a_{d-1,d},a_{dd};S_{d-1,d}}(v,w)/\partial w}$, and
\begin{eqnarray*}
\lefteqn{ F_{a_{dd}|S_{d-1,d}\bigcup a_{d-1,d}}\left(x_{a_{dd}}|\boldsymbol{x}_{S_{d-1,d}}, x_{a_{d-1,d}}\right)} \\
&=& C_{a_{dd}|S_{d-1,d}\bigcup a_{d-1,d}}\left(u_{a_{dd}}|\boldsymbol{u}_{S_{d-1,d}}, u_{a_{d-1,d}}\right) \\
&=& C_{a_{dd}|a_{d-1,d};S_{d-1,d}}\left(C_{a_{dd}|S_{d-1,d}}(u_{a_{dd}}|\boldsymbol{u}_{S_{d-1,d}})|C_{a_{d-1,d}|S_{d-1,d}}(u_{a_{d-1,d}}|\boldsymbol{u}_{S_{d-1,d}})\right),
\end{eqnarray*}
where $C_{a_{dd}|a_{d-1,d};S_{d-1,d}}(v|w) = {\partial C_{a_{d-1,d},a_{dd};S_{d-1,d}}(v,w)/\partial v}$. The terms $C_{a_{d-1,d}|S_{d-1,d}}(u_{a_{d-1,d}}|\boldsymbol{u}_{S_{d-1,d}})$ and $C_{a_{dd}|S_{d-1,d}}(u_{a_{d-1,d}}|\boldsymbol{u}_{S_{d-1,d}})$ can be obtained recursively by taking partial derivatives of the bivariate copulas at previous levels. Given a quantile $q$, the solutions to 
 $$C_{a_{d-1,d}|S_{d-1,d}\bigcup a_{dd}}\left(u_{a_{d-1,d}}|\boldsymbol{u}_{S_{d-1,d}}, u_{a_{dd}}\right) = q, \quad \mbox{and} \quad C_{a_{dd}|S_{d-1,d}\bigcup a_{d-1,d}}\left(u_{a_{dd}}|\boldsymbol{u}_{S_{d-1,d}}, u_{a_{d-1,d}}\right) = q,$$ 
i.e., 
 $$u_{a_{d-1,d}|S_{d-1,d}}^*(q) = C^{-1}_{a_{d-1,d}|S_{d-1,d}\bigcup a_{dd}}\left(q|\boldsymbol{u}_{S_{d-1,d}}, u_{a_{dd}}\right) \quad \mbox{and}$$
 $$u_{a_{dd}|S_{d-1,d}}^*(q) = C^{-1}_{a_{dd}|S_{d-1,d}\bigcup a_{d-1,d}}\left(q|\boldsymbol{u}_{S_{d-1,d}}, u_{a_{d-1,d}}\right),$$ 
are the desired quantiles of a univariate conditional distribution of the copula. With $u_{a_{d-1,d}|S_{d-1,d}}^*(q)$ and $u_{a_{dd}|S_{d-1,d}}^*(q)$, one can further apply the inverse probability integral transform to obtain the predictions $F_{a_{d-1,d}|S_{d-1,d}}^{-1}(q|\boldsymbol{x}_{S_{d-1,d}})$ and $F_{a_{dd}|S_{d-1,d}}^{-1}(q|\boldsymbol{x}_{S_{d-1,d}})$ given the quantile $q$.

For the remaining $d-2$ variables, the conditional distribution can be obtained by a one-dimensional integration. When $j \in S_{d-1,d}$,
\begin{eqnarray*}
\lefteqn{ F_{j|\{(1:d)\setminus j\}}(x_j|\boldsymbol{x}_{\{(1:d)\setminus j\}}) = C_{j|\{(1:d)\setminus j\}}(u_j|\boldsymbol{u}_{\{(1:d)\setminus j\}})
= \int_{0}^{u_j}c_{j|\{(1:d)\setminus j\}}(y|\boldsymbol{u}_{\{(1:d)\setminus j\}})\mathrm{d}y } \\
&=& \frac{\int_{0}^{u_j}c_{1:d}(u_1,\dots,u_{j-1},y,u_{d+1}\dots,u_d)\mathrm{d}y}{c_{\{(1:d)\setminus j\}}(\boldsymbol{u}_{\{(1:d)\setminus j\}})} 
=\frac{\int_{0}^{u_j}c_{1:d}(u_1,\dots,u_{j-1},y,u_{j+1}\dots,u_d)\mathrm{d}y}{\int_{0}^{1}c_{1:d}(u_1,\dots,u_{j-1},y,u_{j+1}\dots,u_d)\mathrm{d}y},
\end{eqnarray*}
where the denominator can be computed using numerical integration methods such as Gaussian quadrature. Given a quantile $q$, the solution to the equation
\[
g(u_j)\equiv \int_{0}^{u_j}c_{1:d}(u_1,\dots,u_{j-1},y,u_{j+1}\dots,u_d)\mathrm{d}y -q\int_{0}^{1}c_{1:d}(u_1,\dots,u_{j-1},y,u_{j+1}\dots,u_d)\mathrm{d}y = 0,
\]
i.e., $u_{j|\{(1:d)\setminus j\}}^*(q) = C^{-1}_{j|\{(1:d)\setminus j\}}(q|\boldsymbol{u}_{\{(1:d)\setminus j\}})$, is the desired quantile of a univariate conditional distribution of the copula. Similarly, one can further apply the inverse probability integral transform to obtain the prediction $F_{j|\{(1:d)\setminus j\}}^{-1}(q|\boldsymbol{x}_{\{(1:d)\setminus j\}})$ given the quantile $q$.

Conditional medians as point estimations can be obtained by setting $q=0.5$. The lower and upper bounds of the $100(1-\alpha)$\% prediction intervals for $0.5<\alpha<1$ can be obtained by setting $q=\alpha/2$ and $q=1-\alpha/2$, respectively. Repeating the above-mentioned procedure for all the variables generates the desired cross prediction results.

\subsection{Simulation from a Conditional Distribution} 
\label{subsec: conditionalsimulation}

An example of a conditional inference of interest is how  extreme values of one variable affect the other variables.
One way to assess this is to simulate from the conditional distribution
$C_{\{(1:d)\setminus j\} |j}(\cdot|q)$
given that one variable has values at an extreme quantile $q$.
Summary statistics from the conditional distribution can indicate which
variables are more affected; this is related to tail dependence.

Simulation of a vine copula is based on the Rosenblatt transform
through a sequence of conditional distributions of the vine copula.
In principle, any permutation of the indices for all variables can be used to simulate data from a vine copula. However, using the order of columns of the vine array usually leads to the simplest computation.
See Sections 6.9.1 and
6.14 of \cite{joe2014dependence}.

Consider an indexing $j_1,\ldots,j_d$ of $d$ variables.
Suppose $p_1,\ldots,p_d$ are pseudo $U(0,1)$ random variables. The
Rosenblatt transform (conditional method) generates $(u_{j_1},\ldots,u_{j_d})$
with a
copula $C$ based on this particular indexing via
$$u_{j_1}=p_1,\ 
u_{j_2}=C_{j_2|j_1}^{-1}(p_2|u_{j_1}),\ 
\ldots,
u_{j_d}=C_{j_d|j_1,\ldots, j_{d-1}}^{-1}(p_d|u_{j_1},\ldots,u_{j_{d-1}}).$$

If the first variable, indexed as $j_1$, is fixed at a given value (say stressed at
upper quantile value $q_1=0.95$), then in the above sequence,
$p_1=q_1$ is fixed instead of being randomly chosen, and $p_2,\ldots,p_d$
are random. These leads to simulation from the conditional
distribution $C_{j_2,\ldots,j_d|j_1}(\cdot|q)$.

In terms of the implementation via a vine array, the representation via
a vine array needs to be modified so that variable $j_1$ is in column 1
of the vine array.
This is always possible and then column 2 will have a
variable that is paired in tree 1 with this specific variable.
The algorithm is then a small modification of Algorithm 17 in
\cite{joe2014dependence}.

The ideas extend to simulation from the conditional distribution
$C_{j_{m+1},\ldots,j_d|j_1\cdots j_m}(\cdot|q_{j_1},\ldots,q_{j_m})$.
with $1<m<d$.
The algorithm based on a vine array can be readily modified if
the variables index by $j_{1},\ldots,j_m$ form a marginal vine copula
and the order $j_1,\ldots,j_m$ can appear in the first $m$ columns
of a vine array representation of the vine.
In this case, all of the conditional distributions in the sequence
can be obtained sequentially and recursively based on the pair-copulas in the vine.
It is a numerical issue to decide which other conditional distributions
can be easily simulated with a given vine copula.

\section{Comparison of Cross Predictions} \label{sec: sim}

Our intention for analyzing the financial dataset introduced in Section \ref{sec: data}
is to show that cross predictions based on vine copulas can be better
than those based on Gaussian copulas or classical multiple regression.
Because vine copulas are more flexible to allow for nonlinear
prediction quantiles and heteroscedastic prediction intervals,
comparisons of cross predictions are made based on several
criteria introduced in Section~\ref{subsec: scores}.
Then we summarize a simulation study to show when
the method of vine copulas is indeed better in Section~\ref{subsec: sim}.

\subsection{Comparison criteria} \label{subsec: scores}

Cross prediction methods can be evaluated and compared
based on mean absolute error of prediction, root mean squared error of prediction,
and interval score of prediction intervals.
A reference for interval score (IS) is \cite{gneiting2007strictly}.
The evaluation methods can be summarized separately for the prediction of each variables and/or
as averages over all $d$ variables.

A training set is used to fit different models and a holdout or test
set of size $n_{\text{test}}$ is used to 
compare the models.
The measures given below have the form of averages over $d$ variables.

\begin{itemize}
	\itemsep=0pt
	\item The mean absolute error (MAE) measures a model's performance on point
	estimations:
	\[
	\text{MAE}(\mathcal{M}) =\frac{1}{d} \sum_{j = 1}^{d} \left\{\frac{1}{n_{\text{test}}}\sum_{i=1}^{n_\text{test}}\left|\tilde{x}_{ij}-\widehat{\tilde{x}}^{\mathcal{M}}_{ij}\right|\right\},
	\]
	where $\tilde{x}_{ij}$ is the observation for the $j$th variable of the $i$th sample in the test set and $\widehat{\tilde{x}}^{\mathcal{M}}_{ij}$ is the predicted conditional expectation or conditional median for that observation based on a fitted model $\mathcal{M}$.

	\item The root mean squared error (RMSE) measures a model's performance on point
	estimations:
	\[
	\text{RMSE}(\mathcal{M}) = \frac{1}{d}\sum_{j = 1}^{d}\sqrt{\frac{1}{n_{\text{test}}}\sum_{i=1}^{n_\text{test}}\left(\tilde{x}_{ij}-\widehat{\tilde{x}}^{\mathcal{M}}_{ij}\right)^2} \,.
	\]

	\item The interval score (IS) is a scoring rule for quantile and interval predictions. In the case of the central $100(1-\alpha)$\% prediction interval, let $\widehat{l}^{\mathcal{M}}_{ij}$ and $\widehat{u}^{\mathcal{M}}_{ij}$ be the predicted quantiles at levels $\alpha/2$ and $1-\alpha/2$ by a fitted model $\mathcal{M}$ for the $j$th variable of the $i$th sample in the test set. The interval score for cross prediction is defined as
	\[
	\text{IS}(\mathcal{M}) = \frac{1}{d} \sum_{j = 1}^{d} \Biggl\{
	\frac{1}{n_{\text{test}}}\sum_{i=1}^{n_\text{test}}\Bigl[\bigl(\widehat{u}^{\mathcal{M}}_{ij}-\widehat{l}^{\mathcal{M}}_{ij} \bigr) + \frac{2}{\alpha}\bigl(\widehat{l}^{\mathcal{M}}_{ij} - \tilde{x}_{ij}\bigr)\mathbb{I}\{\tilde{x}_{ij} < \widehat{l}^{\mathcal{M}}_{ij}\} + \frac{2}{\alpha}\bigl(\tilde{x}_{ij} - \widehat{u}^{\mathcal{M}}_{ij}\bigr)\mathbb{I}\{\tilde{x}_{ij} > \widehat{u}^{\mathcal{M}}_{ij}\}\Bigr] \Biggr\}.
	\]
Smaller interval scores imply superior prediction performance
that has prediction intervals that are not too long and do not miss the
``true value'' by much when the prediction interval does not contain
the true value in the test set.
\end{itemize}

\subsection{Simulation Studies} \label{subsec: sim}

In this section, the effectiveness of the proposed cross prediction method based on vine copulas is demonstrated on simulated datasets. There are five variables in each simulated dataset. 
The vine copula structure is chosen to match the behavior of the fitted
vine copula for the dataset discussed in Section \ref{sec: data}.
The variables are generated from the following univariate marginal distributions and multivariate copula models:
\begin{enumerate}
	\item Each variable follows a univariate normal $(0,1)$ distribution. The vine array, vine bivariate copula family matrix, and vine bivariate copula parameter matrix are as follows:
	\[
	A = 
	\begin{bmatrix}
	1 & 1 & 2 & 2 & 3\\
	& 2 & 1 & 3 & 2\\
	&   & 3 & 1 & 4\\
	&   &   & 4 & 1\\
	&   &   &   & 5\\
	\end{bmatrix},
	F_1 =
	\begin{bmatrix}
	- & \text{N} & \text{N} & \text{N} & \text{N}\\
	& - & \text{N} & \text{N} & \text{N}\\
	& & - & \text{N} & \text{N}\\
	& & & - & \text{N}\\
	& & & & - \\
	\end{bmatrix},
	P_1 =
	\begin{bmatrix}
	- & 0.8 & 0.6 & 0.5 & 0.7\\
	& - & 0.4 & 0.5 & 0.3\\
	& & - & 0.3 & 0.2\\
	& & & - & 0.1\\
	& & & & - \\
	\end{bmatrix},
	\]
	where N in the copula family matrix stands for bivariate Gaussian (normal) copula. This is equivalent to a multivariate Gaussian distribution. Note that this simulation scenario has pairs of variables with strong dependence in tree 1.
	\item Each variable follows a univariate log-normal $(0,1)$ distribution. The vine array, vine bivariate copula family matrix, and vine bivariate copula parameter matrix are the same as the first case.
	\item Each variable follows a univariate normal $(0,1)$ distribution. The vine array, vine bivariate copula family matrix, and vine bivariate copula parameter matrix are as follows:
	\[
       \arraycolsep 3pt
	A = 
	\begin{bmatrix}
	1 & 1 & 2 & 2 & 3\\
	& 2 & 1 & 3 & 2\\
	&   & 3 & 1 & 4\\
	&   &   & 4 & 1\\
	&   &   &   & 5\\
	\end{bmatrix},
	F_2 =
	\begin{bmatrix}
	- & \text{t} & \text{t} & \text{BB1} & \text{BB1}\\
	& - & \text{F} & \text{F} & \text{N}\\
	& & - & \text{G} & \text{N}\\
	& & & - & \text{F}\\
	& & & & - \\
	\end{bmatrix},
	P_2 =
	\begin{bmatrix}
	- & 0.7 (5) & 0.8 (4) & 1 (2) & 2 (1.5)\\
	& - & 3 & 2 & 0.4\\
	& & - & 1.2 & 0.2\\
	& & & - & 1.2\\
	& & & & - \\
	\end{bmatrix},
	\]
	where N, t, G, F in the copula family matrix stand for bivariate Gaussian, Student's t, Gumbel, and Frank copulas, respectively. Note that the Student's t and BB1 copula families have two parameters while all other copula families have one parameter. This model adds more tail dependence than the multivariate Gaussian distribution. Note that upper or lower tail dependence for all bivariate copulas in the first tree implies upper or lower tail dependence for all pairs of variables (\cite{joe2010tail}). This result implies that if tail dependence is seen in all bivariate plots,
	then the first tree of the vine copula should have bivariate copulas with
	tail dependence.
	\item Each variable follows a univariate log-normal $(0,1)$ distribution. The vine array, vine bivariate copula family matrix, and vine bivariate copula parameter matrix are the same as the third case.
\end{enumerate}

Samples of size 1000 are generated for each of the four cases. Among each sample, 800 observations are randomly selected as the training set and another 200 observations are treated as the test set. Cross prediction using four methods is considered in the simulation study: (1). linear regression, (2). linear regression with logarithmic transformation of all the variables (for cases 2 and 4 only), (3). Gaussian copula prediction, and (4). vine copula prediction. For linear regression with or without logarithmic transformation, five separate linear models are fitted for each variable. For copula prediction methods, one model for the joint distribution of the five variables is fitted, the five conditional distributions are derived from the joint distribution. The Gaussian copula prediction can be considered as a special case of the vine copula prediction, where the bivariate copula models on the vine edges are all Gaussian. Different methods are trained on the training set and used to obtain the conditional medians (for copula models) or conditional expectations (for linear regression models) as point predictions as well as $100(1-\alpha)$\% prediction intervals on the test set. For the vine copula regression, the candidate bivariate copula families include 
Gaussian, Student's t, Clayton, Gumbel, Frank, and BB1 copulas, 
as well as their corresponding survival and reflected counterparts. 
This covers a range of different tail asymmetries and strength of dependence in the joint tails. 
The copula with reflected first variable, survival copula, and copula with reflected second variable, based on a given bivariate copula $C(u,v)$, are defined as
$$ C_u(u,v) = v-C(1-u, v), \ 
 C_s(u,v) = u+v-1+ C(1-u, 1-v) \ 
\mbox{and} \ 
 C_v(u,v) = u-C(u, 1-v),$$
respectively. Vine structure learning and bivariate copula selections are performed using the guidelines described in Sections~\ref{subsec: vine_struct}
and \ref{subsec: biv_cop_select}. The simulation study is repeated 100 times for each case using each method. The averages of the comparison criteria and their standard deviations are reported in Table~\ref{tab: sim_performance}. 

\begin{table}[!ht]
	\centering
	\begin{tabular}{*{5}{l}}
		\toprule
		Case & Method & MAE & RMSE & 80\% IS \\\midrule
		\multirow{3}{*}{1} & Linear regression & 0.536 (0.032) & 0.680 (0.042) & 2.406 (0.148) \\
		& Gaussian copula & 0.536 (0.033) & 0.679 (0.042) & 2.404 (0.149) \\
		& Vine copula & 0.523 (0.032) & 0.670 (0.042) & 2.318 (0.146) \\\midrule
		\multirow{4}{*}{2} & Linear regression & 0.787 (0.079) & 1.320 (0.225) & 4.837 (0.602) \\
		& Regression with log-transform & 0.766 (0.074) & 1.421 (0.268) & 3.628 (0.350) \\
		& Gaussian copula & 0.766 (0.075) & 1.427 (0.293) & 3.630 (0.370) \\
		& Vine copula & 0.733 (0.071) & 1.298 (0.243) & 3.403 (0.361) \\\midrule
		\multirow{3}{*}{3} & Linear regression & 0.376 (0.016) & 0.483 (0.021) & 1.726 (0.075)\\
		& Gaussian copula & 0.375 (0.016) & 0.482 (0.020) & 1.724 (0.075) \\
		& Vine copula & 0.370 (0.016) & 0.481 (0.022) & 1.671 (0.074) \\\midrule
		\multirow{4}{*}{4} & Linear regression & 0.611 (0.076) & 1.073 (0.219) & 3.942 (0.622) \\
		& Regression with log-transform & 0.583 (0.065) & 1.111 (0.212) & 2.815 (0.341) \\
		& Gaussian copula & 0.584 (0.066) & 1.119 (0.229) & 2.818 (0.345) \\
		& Vine copula & 0.567 (0.069) & 1.062 (0.221) & 2.681 (0.360) \\
		\bottomrule
	\end{tabular}
	\caption{The mean absolute error (MAE), root mean squared error (RMSE), and interval score (IS) of the simulation test performance for different cases comparing different methods. The numbers in parentheses are the corresponding standard deviations over 100 repetitions.
	The numbers of times that IS for vine copula is smaller than IS for regression (cases 1,3) and smaller than for regression with log-transform (cases 2,4) are 94, 96, 98, and 93 out of the 100 repetitions for cases 1, 2, 3, and 4, respectively.
}
	\label{tab: sim_performance}
\end{table}

\begin{figure}[!ht] 
	\centering
	\includegraphics[width=\linewidth]{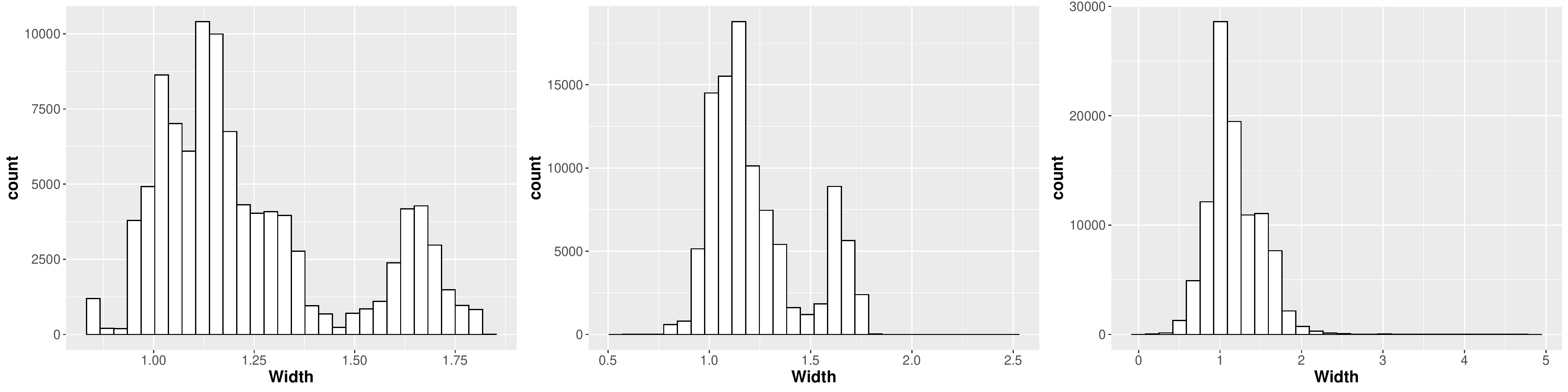}
	\caption{The histograms of the widths of the 80\% prediction intervals generated by linear regression, Gaussian copula cross prediction, and vine copula cross prediction for simulation case 3 based on all the 100 repetitions. The total number of data points in each histogram in $5 \text{ (variables)} \times 200 \text{ (number of test cases)} \times 100 \text{ (repetitions)} = 100,000$.}
	\label{fig: width_sim}
\end{figure}

From Table~\ref{tab: sim_performance}, it can be seen that in cases 1 and 3 where the univariate distributions are Gaussian, cross prediction with Gaussian copula has similar performance to linear regression. In cases 2 and 4 where the univariate distributions are log-normal, the performance of cross prediction with Gaussian copula is similar to that of linear regression with log-transformation. Overall, in all the four simulation scenarios, cross prediction with vine copula works the best in terms of point and interval prediction performance. Lower IS values indicate that vine copulas work better in generating more informative prediction intervals of diverse widths than linear regression. For example, the histograms of the widths of the 80\% prediction intervals generated by different methods for simulation case 3 based on all the 100 repetitions are shown in Figure~\ref{fig: width_sim}. The residual standard deviations by linear regression for the five variables in simulation case 3 are 0.614, 0.450, 0.438, 0.322, and 0.382, respectively. The standard deviation in predicting the first variable is larger than the other four variables. Therefore, the histogram is bimodal since not all variables are predicted to the same accuracy.

\section{Credit Spread Dataset of Chinese Corporate Bonds} \label{sec: data}

The financial dataset used for conditional inference and copula modeling contains the sector-wise mean daily credit spreads of the AAA-rated corporate bonds from 24 sectors traded on the Chinese bond market. All the data are collected and provided by the KYZ project (\url{http://kyz.pingan.com/}) at Ping An Asset Management. The 24 sectors contained in the dataset are: catering \& tourism, city investment, real estate, iron, transportation, public utilities, chemical engineering, construction materials, construction, coal, motor manufacturing, commerce \& retail, oil \& gas, food \& beverage, culture \& media, nonferrous metals, equipment manufacturing, electronics, electricity generation, electricity supply, national railway, health care, comprehensive finance, and other comprehensive industries. The credit spread data of these 24 sectors were collected on a daily basis from August 5, 2013 to December 31, 2016 with a sample size of 852. A difference at lag 1 is taken, thus the analysis focuses on the series of daily changes of the sector-wise mean credit spreads. Data from the first 600 days are used as the training set, while data from the last 252 days are left for out-of-sample testing.

For the credit spread dataset of Chinese corporate bonds, it is of primary interest to model the underlying dependence structure among sectors. As a flexible tool in fitting high-dimensional distributions, vine copulas with tail dependence can be used for multivariate dependence modeling. After the multivariate joint distribution is fitted,  conditional inferences can be performed:  cross predictions to predict one sector given all other sectors, or risk transfer to analyze the conditional distribution of all other sectors given observations from one sector being extreme.

As discussed in Section~\ref{sec: vine_cop}, analysis with copula models usually involves two steps: the first step identifies and fits the univariate marginal distributions, and the second step fits appropriate copula families to represent the multivariate dependence structure. This section summarizes the preprocessing procedure transforming the credit spread data to uniform random variables. Specifically, the decomposition of bond yields is discussed in Section~\ref{subsec: yield}. Section~\ref{subsec: data_collection} elaborates on the credit spread data collection procedures. Section~\ref{subsec: filter} introduces the ARMA-GARCH filter to remove the temporal dependence in the dataset. Procedures to smooth sector-specific outliers are presented in Section~\ref{subsec: outlier}.

\subsection{Decomposition of Bond Yields} \label{subsec: yield}

The return rate of a bond is usually measured by yield to maturity (YTM, or simply yield when it does not cause any ambiguity), which is the annualized rate of return earned by an investor who buys the bond today at the market price, assuming that the bond is held until maturity, and that all coupon and principal payments are made on schedule.
The yield of a bond can be decomposed into three additive parts as follows.
\begin{enumerate}
	\item The {\em risk-free rate}, which is usually approximated by the return rate of the Treasury bills in the United States or AAA-rated government bonds in other countries;
	\item The {\em risk premium}, which is also known as the {\em credit spread}. This is the extra return investors expect to get in order to compensate for the potential credit risk (e.g., losses caused by bond default or bond downgrade), and can be regarded as the credit quality of the bond issuer recognized by the market. The credit spread of a bond can come from several sources: ratings of the bond, the sector which the bond issuer corporate belongs to, and bond-specific variations. By taking sector-wise means, the effects of bond-specific variations are eliminated. Therefore, the analysis here focuses on the sector-specific credit spreads.
	\item The {\em liquidity premium}. This is the extra return to compensate for the potential loss due to lack of liquidity. The liquidity premium of a bond is usually a fairly small number, especially for AAA-rated bonds. Therefore, the liquidity premium is often omitted in credit analysis.
\end{enumerate}

\subsection{Data Collection and Processing Procedures} \label{subsec: data_collection}

One of the important characteristics of bonds is that intrinsically similar bonds (bonds with same ratings) may have different maturities. Therefore, one can create a curve, namely the yield curve, to model the yields across different maturities for a set of similar bonds. The credit spread of a bond can thus be calculated based on the nearest benchmark on the ChinaBond corporate bond yield curves (available on \url{https://yield.chinabond.com.cn/}) matching the rating and maturity of that bond. The raw dataset of the sector-wise mean daily credit spreads of the AAA-rated corporate bonds from 24 sectors is collected and prepared by Ping An Asset Management according to the following four steps.
\begin{enumerate}
\item \textit{Sample inclusion}. All the available medium-term bonds on the Chinese bond market with maturity from one to five years are selected starting from 2013. Bonds with special terms such as perpetual bonds, bonds with guarantee, and installment bonds are not selected. Bond yield at the exercise time is used for bonds with embedded options. Bond samples with public ratings at AAA, AA+ or AA on ChinaBond.com are considered for each sector.

\item \textit{Sample exclusion}. Bonds that were newly issued within 90 days are excluded from the sample pool to remove possible yield fluctuations when they were initially listed on the market. If the public rating of a bond is changed to any level lower than AA on any given day, this bond is permanently removed from all sample pools starting from that day. Otherwise, this bond is temporarily removed from its original sample pool and will be included to another sample pool again after 90 days if its public rating stays the same during this period. For example, if the rating of a bond was AAA before day 100 and AA starting from day 101, this bond is included in the sample pool for AAA-rated bonds from day 1 to 100, temporarily removed from all sample pools from day 101 to 190, and is included in the sample pool for AA-rated bonds from day 191 onwards. Only the sample pool for AAA-rated bonds is used for the study in this paper.

\item \textit{Initial outlier processing}. At this step, bonds with unusual yield changes are removed, such that samples included in the pool are as representative as possible for each sector. In particular, bond samples that satisfy the following two conditions are removed:
\begin{itemize}
	\item A bond sample $b_i$ is removed from the pool if $v_t(b_i) - v_{t-1}(b_i) > 0.20\%$ and $y(r_i(t), m_i(t)) - y(r_i(t-1), m_i(t-1)) < 0.15\%$, where $v_t(b_i)$ is the yield of bond $b_i$ given by ChinaBond.com on day $t$ and $y(r_i(t), m_i(t))$ is the nearest benchmark yield of bonds with rating $r_i(t)$ and maturity $m_i(t)$ according to the ChinaBond corporate bond yield curve. However, if this condition holds for more than 67\% of the bond samples from the same sector, these bond samples are not removed.
	\item A bond sample $b_i$ is removed from the pool if $v_t(b_i) > 9\%$ on any day. The exceptionally high yield usually implies that the bond is too risky to purchase.
\end{itemize}
Bond samples removed according to the previous two conditions can be included back to the pool if their yields appear normal again when $v_t(b_i) < v_{t_0}(b_i) + 0.05\%$, where $v_{t_0}(b_i)$ is the yield of the bond $b_i$ on the day before it was removed from sample pool.

\item \textit{Sector-wise spread credit calculation}. For each bond $b_i$ remaining in the sample pool, its credit spread on day $t$, denoted by $s_t(b_i)$, is calculated as $s_t(b_i) = v_t(b_i) - y(r_i(t), m_i(t))$. 
The sector-wise mean credit spread is given by the arithmetic average of $s_t(b_i)$ for all the selected bond samples from a given sector on that day. This can be treated as the expected default risk of that sector as a whole recognized by the bond market.  
\end{enumerate}

\subsection{ARMA-GARCH Filter} \label{subsec: filter}

For financial time series, an ARMA-GARCH filter with innovation following skewed Student's t distribution is usually applied to make the data independent and identically distributed.

For the Chinese corporate bond dataset, the ARMA(1,1)-GARCH(1,1) filter is applied to remove the serial dependence. Denote the observation of sector $j$ at time $t$ by $X_{t,j}$. The ARMA(1,1)-GARCH(1,1) model with skewed Student's t innovations can be expressed as:
\begin{align*}
\begin{cases}
X_{t,j} = \mu + \phi X_{t-1,j} + \epsilon_{t,j} + \theta\epsilon_{t-1,j}, \\
\epsilon_{t,j} = Z_{t,j}\sigma_{t,j}, \\
\sigma^2_{t,j} = \omega + \alpha\epsilon^2_{t-1,j} + \beta\sigma^2_{t-1,j},
\end{cases} 
\end{align*}
where $t = 1, \dots, T$, $j = 1, \dots, d$, and the innovations $Z_{t,j}$ follow skewed Student's t distribution. The ARCH test and Ljung-Box test are applied to the residuals; test results indicate that the ARMA(1,1)-GARCH(1,1) model works well in removing the serial dependence for most of the sectors. Specifically, catering \& tourism has significant $p$-values for ARCH and Ljung-Box tests. However, catering \& tourism is removed in the subsequent analysis due to low correlation with other sectors. Iron, chemical engineering, coal, nonferrous metals, and electricity generation have significant $p$-values for the Ljung-Box test. These sectors have more outliers than average, which could lead to the significant $p$-values. In comparison, if the GARCH(1,1) filter is applied to the Chinese corporate bond data, it fails to converge for a few sectors.

From the Pearson's correlation matrix of the filtered residuals by
sector,
most of the sectors are moderately correlated with each other. Nevertheless, there are a few sectors whose residuals almost have no correlation with other sectors. 
We therefore decided to
remove those sectors with weak correlation before 
fitting vine dependence models.
Table~\ref*{tab: abs_cor} displays the mean of the absolute Pearson's correlation coefficients between the residuals of each sector with those for all other sectors. There are four sectors, catering \& tourism, food \& beverage, health care, and comprehensive finance, whose residuals barely correlate with any other sector. The mean of the absolute values of their correlation coefficients with other sectors are also less than 0.1. Therefore, these four sectors are removed from the analysis and the remaining 20 sectors are kept for the 
further dependence structure study.


\begin{table}[!ht]
	\footnotesize
	\centering
	\begin{tabular}{*{4}{c}}
		\toprule
		catering \& tourism & city investment & real estate & iron \\
		0.089 & 0.250 & 0.244 & 0.180 \\\midrule
		transportation & public utilities & chemical engineering & construction materials \\ 
		0.290 & 0.214 & 0.156 & 0.221 \\\midrule
		construction & coal & motor manufacturing & commerce \& retail \\
		0.262 & 0.238 & 0.170 & 0.215 \\\midrule
		oil \& gas & food \& beverage & culture \& media & nonferrous metals \\
		0.161 & 0.067 & 0.179 & 0.230 \\\midrule
		equipment manufacturing & electronics & electricity generation & electricity supply \\
		0.158 & 0.187 & 0.228 & 0.128 \\\midrule
		national railway & health care & comprehensive finance & other comprehensive industries \\
		0.109 & 0.059 & 0.062 & 0.179 \\
		\bottomrule
	\end{tabular}
	\caption{The mean of the absolute Pearson's correlation coefficients of the residuals for each sector with all other sectors.}
	\label{tab: abs_cor}
\end{table}

\subsection{Outlier Identification and Processing} \label{subsec: outlier}

\begin{figure}[!ht] 
	\centering
	\includegraphics[width=\linewidth]{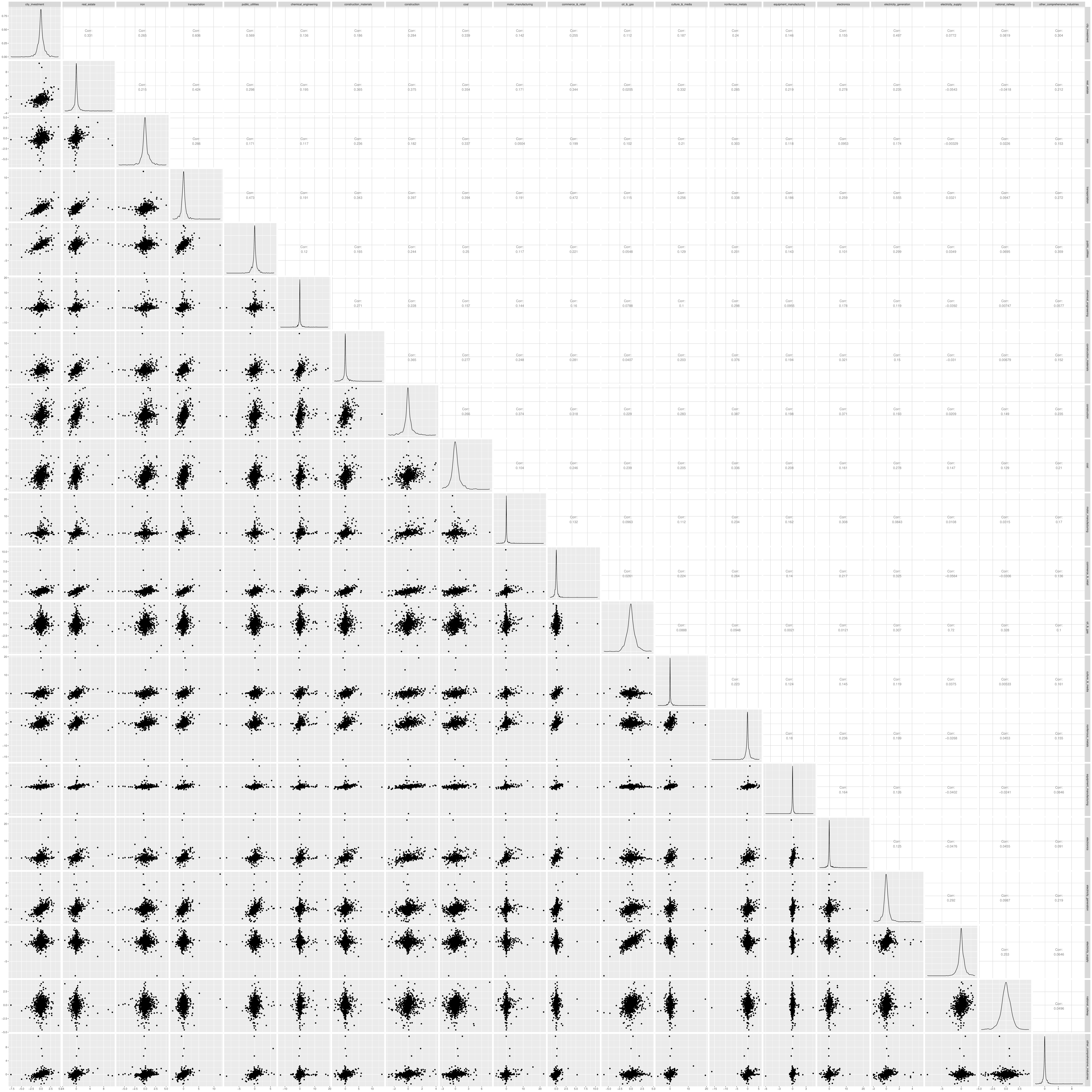}
	\caption{The matrix of the pairwise scatter plots of the residuals of the remaining 20 sectors.} \label{fig: scatter_plot_matrix}
\end{figure}

Figure~\ref{fig: scatter_plot_matrix} shows the matrix of pairwise scatter plots of residuals from the remaining 20 sectors. This serves as a diagnostic check before fitting a dependence model. From the figure, it can be seen that for some sectors, the pairwise scatter plots show a cross star pattern rather than an elliptical shape. When there are a large number of points lying close to the $x$- and $y$-axes on the scatter plots of one sector versus others, this phenomenon suggests that the large deviations of these points cannot be well explained by the residuals from other sectors. These values may be caused by sector-specific variations and do not contribute much to the analysis of the underlying dependence structure among all sectors. Particularly, the initial outlier processing step in Section~\ref{subsec: data_collection} 
can lead to some non-smoothness when the set of bonds used in an average
varies by day (i.e., when bonds are removed or added to the sample pool). It cannot remove sector-specific large variations, either. Therefore, it is necessary to identify and smooth these data points.

To identify the abnormal points, we define a point to be an outlier for a sector on a particular day if the residual for this sector is more than 3 standard deviations away from the mean while the residuals of all other sectors are within 2 standard deviations from their means on the same day. On average, 3.15 out of 852 observations are identified as outliers for each sector. After identifying all outliers, smoothing is performed on the lag 1 difference series. In particular, the outlier as well as the two observations on the previous and next days are smoothed. Denote the mean of these three observations by $m$. For each outlier, two normal random noises with mean 0 and standard deviation equal to 0.1 times the standard deviation of the difference series of that sector are generated; denote the two noises by $e_1$ and $e_2$. These three points are then smoothed to $m - e_1$, $m + e_1 + e_2$, and $m - e_2$, respectively. This ensures that the original time series can be recovered if the cumulative sums are taken on the smoothed difference series. Since the standard deviations of the generated random noises are very small, changing the noise generation seed does not affect the subsequent dependence analysis such as the estimated vine structure of the copula model.

\begin{figure}[!ht] 
	\centering
	\includegraphics[width=.8\linewidth]{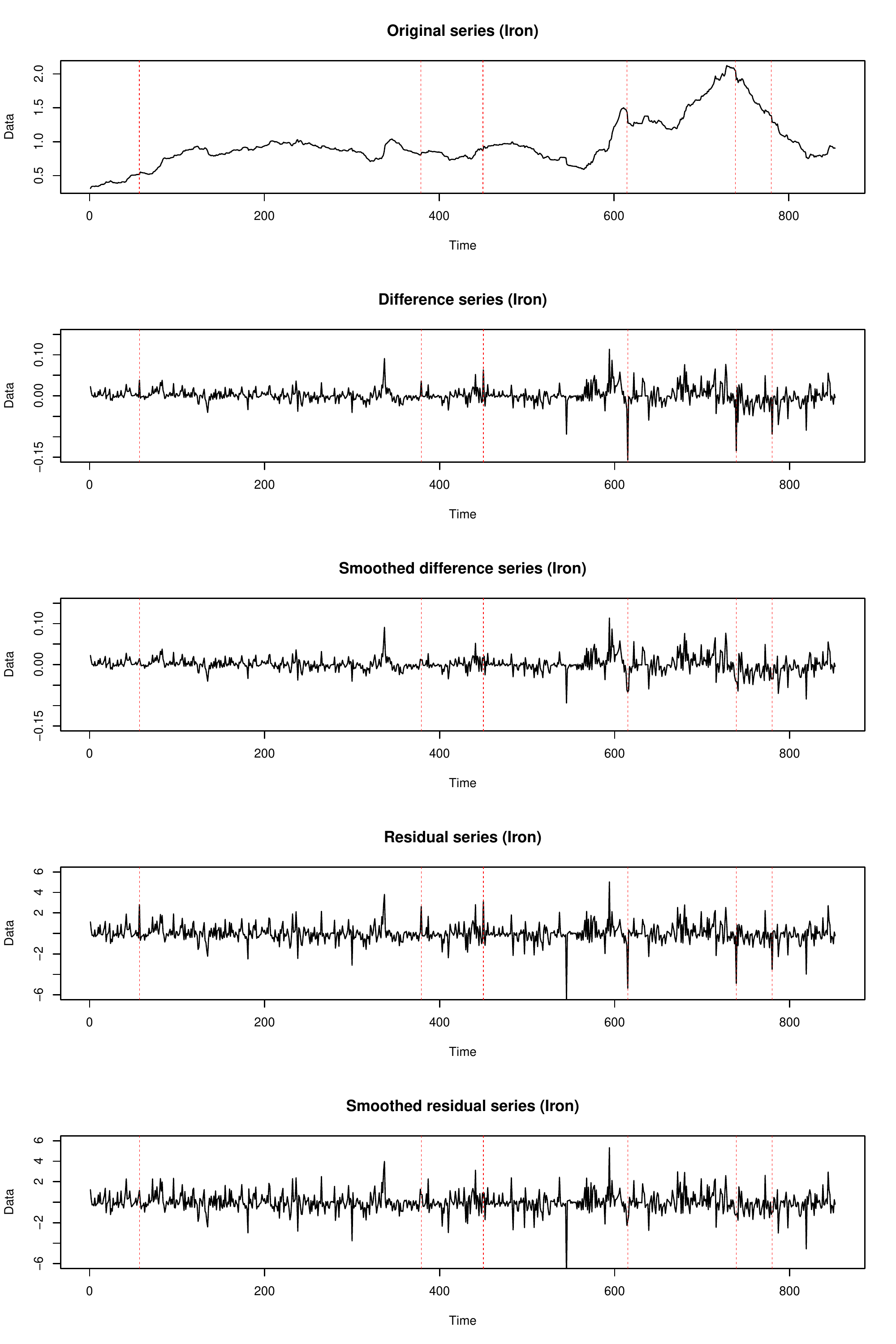}
	\caption{The plots of the original series, difference series, difference series after smoothing, original residual series, and residual series after smoothing of the iron sector. Red vertical lines indicate the locations of the outliers.
} 
\label{fig: smoothed}
\end{figure}

The ARMA(1,1)-GARCH(1,1) filter is reapplied to the smoothed series to obtain the new residual series. Figure~\ref{fig: smoothed} shows the plots of the original series, difference series, difference series after smoothing, original residual series, and residual series after smoothing of the iron sector. It can be seen smoothing makes the changes of the residuals series less abrupt on the outlier days while leaving the residuals on the remaining days mostly unaffected.

After obtaining the new residual series, the skewed Student's t distributions are fitted as the univariate distribution models. The probability integral transform is taken to transform the residual series to u-scores and make sure the transformed data follow the Uniform (0,1) distribution. The u-scores are then used for fitting copula models.

\section{Applications to the Credit Spread Dataset} \label{sec: app}

In this section, copula models are fitted on the residual series of the credit spread dataset of Chinese corporate bonds and the proposed conditional inference method is also applied to this dataset. The dependence structure across sectors suggested by copula models is elaborated in Section~\ref{subsec: dependence_struct}. The performance of the proposed cross prediction method based on vine copulas is evaluated in Section~\ref{subsec: data_performance}. 
The assessment of risk transfer conditioned on the extreme quantile of a central sector is presented in Section~\ref{subsec: risk}.

\subsection{Examining the Dependence Structure Among Sectors} \label{subsec: dependence_struct}

When analyzing financial data collected across multiple sectors, it is often of interest to examine the underlying dependence structure among different sectors. For example, \cite{agca2017credit} study how credit risk propagates through sectors across multiple tiers of supply chains. 

As a graphical dependence model, vine, especially the trees at the first few levels, provides insights on the internal dependence and conditional dependence structure among the variables. In the first tree of a vine, variables with the strongest correlations are connected with the symmetric difference constraint satisfied. Therefore, variables that are a short distance away in the first tree should also have stronger dependence than those that are a long distance away.

\begin{figure}[!ht] 
	\centering
	\includegraphics[width=\linewidth]{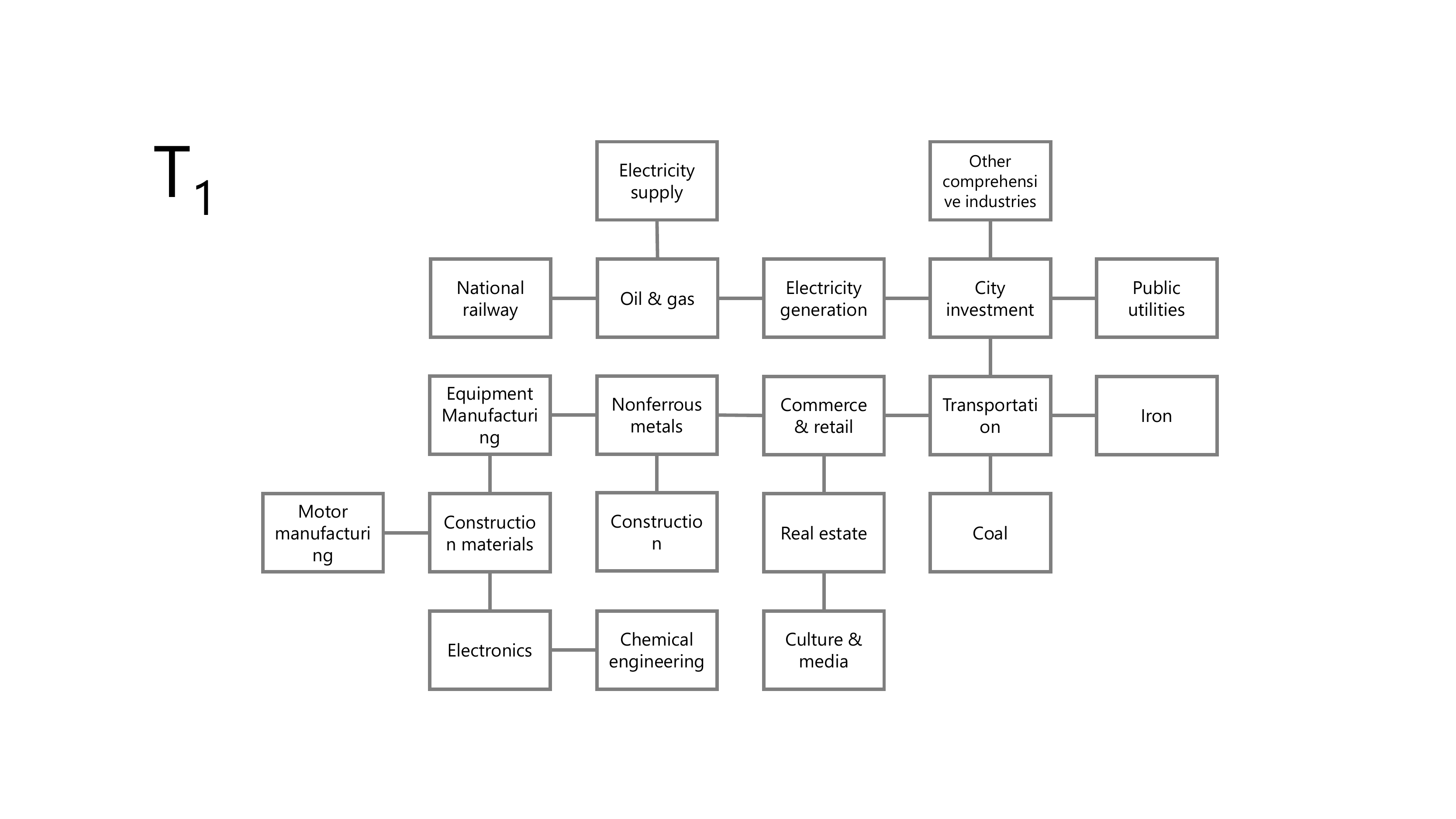}
	\includegraphics[width=\linewidth]{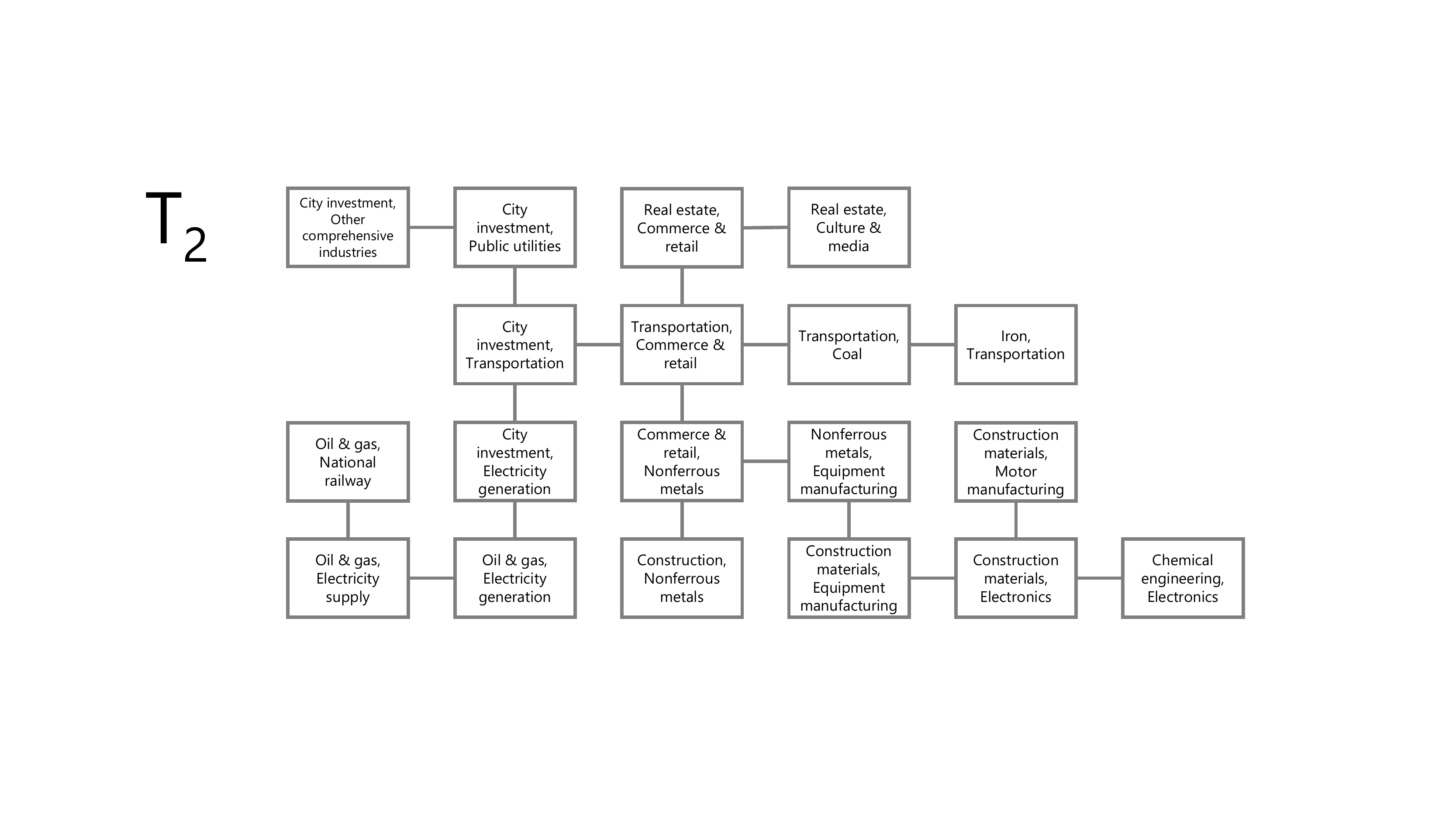}
	\caption{Visualization of the first two levels of the R-vine array.
	} \label{fig: tree_structure}
\end{figure}

The vine array for the fitted R-vine copulas for the training set of the filtered residual series of the credit spread dataset will be provided as
supplementary material.
A visualization of the first two trees of the fitted R-vine is provided in Figure~\ref{fig: tree_structure}. From the first tree, it can be seen that sectors such as nonferrous metals, commerce \& retail, transportation, and city investment are placed at the central positions in the graph. To analyze their association with other sectors, transportation is taken as an example. Based on $T_1$, the distance-1 neighbors of transportation are commerce \& retail, coal, iron, and city investment. The distance-2 neighbors of transportation are nonferrous metals, real estate, electricity generation, public utilities, and other comprehensive industries. The distance-3 neighbors of transportation are equipment manufacturing, construction, culture \& media, and oil \& gas.

The correlation of the normal scores $\rho_N$, the lower and upper semi-correlations  $\rho_N^-$ and $\rho_N^+$ (which are the correlations of the observations in the lower and upper quadrants, respectively) are metrics to measure the strength of the dependence of bivariate copulas. To measure the tail dependence of bivariate copulas, the tail-weighted dependence measures $\zeta_\alpha^U$ and $\zeta_\alpha^L$ in \cite{lee2018tail} can be used as the metrics. For $\alpha>0$, the upper-tail-weighted dependence measure $\zeta_\alpha^U$ for a bivariate copula $C(u_1, u_2)$ is defined as
\[
\zeta_\alpha^U(C) = 2 - \vartheta_\alpha,
\]
where $\vartheta_\alpha$ and $\upsilon_\alpha$ are defined as
\begin{equation}
\vartheta_\alpha = \frac{\alpha + \alpha(1+\alpha)\upsilon_\alpha}{\alpha-(1+\alpha)\upsilon_\alpha}
\quad \mbox{and} \quad
\upsilon_\alpha = \frac{1}{2}\mathbb{E}\left(\left|U_1^\alpha - U_2^\alpha\right|\right), \label{eq: upsilon}
\end{equation}
respectively, with $(U_1,U_2)\sim C$. The lower-tail-weighted dependence measure $\zeta_\alpha^L$ can be defined in a similar manner with $U_1$ and $U_2$ in Equation~\eqref{eq: upsilon} replaced by $1-U_1$ and $1-U_2$. Larger $\zeta_\alpha^U$ and $\zeta_\alpha^L$ indicate stronger tail dependence.

The means of $\rho_N$, $\rho_N^-$, $\rho_N^+$, $\zeta_{10}^U$, and $\zeta_{10}^L$ across all the distance-1, 2, and 3 neighbors of transportation are summarized in Table~\ref{tab: dependence_metrics}. It can be seen that all the dependence measures decrease as the distance between a sector and the transportation sector increases. This suggests that vine is an interpretable tool in visualizing the underlying dependence structure among variables.

\begin{table}[!ht]
	\centering
	\begin{tabular}{*{4}{l}}
		\toprule
		Metrics & Distance-1 & Distance-2 & Distance-3 \\\midrule
		$\rho_N$ & 0.505 & 0.440 & 0.318 \\
		$\rho_N^-$ & 0.412 & 0.383 & 0.283 \\
		$\rho_N^+$ & 0.497 & 0.367 & 0.334 \\
		$\zeta_{10}^U$ & 0.328 & 0.247 & 0.162 \\
		$\zeta_{10}^L$ & 0.321 & 0.269 & 0.134 \\
		\bottomrule
	\end{tabular}
	\caption{The means of $\rho_N$, $\rho_N^-$, $\rho_N^+$, $\zeta_{10}^U$, and $\zeta_{10}^L$ across all the distance-1, 2, and 3 neighbors of transportation.}
	\label{tab: dependence_metrics}
\end{table}

\subsection{Evaluation of the Cross Prediction Performance} \label{subsec: data_performance}

This section evaluates the effectiveness of the proposed cross prediction method based on vine copulas. Different ways of deciding on the vine structure
in Section \ref{subsec: vine_struct} led to similar comparison results.

Similar to the simulation studies in Section~\ref{subsec: sim}, the performance of the following three cross prediction methods are considered: (1). linear regression, (2). Gaussian copula prediction, (3). vine copula prediction. The candidate bivariate copulas for vine copula cross prediction include Gaussian, Student's t, Clayton, Gumbel, Frank, BB1, and BB8 copulas, as well as their corresponding survival and reflected counterparts.

Copula and linear regression models are fitted on the training set, and the test set is used for the purpose of performance evaluation. The prediction targets are the ARCH-GARCH filtered residuals of the 20 sectors. All univariate distributions are modeled by skewed Student's t distribution. The conditional expectations (or medians) and the 80\% prediction intervals are obtained for each method. For copula predictions, the lower and upper bounds of the 80\% prediction interval are the conditional 10\% and 90\% quantiles, respectively. The absolute error (MAE), root mean squared error (RMSE), and interval score (IS) comparing different methods in predicting the ARMA-GARCH filtered residuals after smoothing for different sectors in the credit spread dataset of Chinese corporate bonds are reported in Table~\ref{tab: data_performance}. For sector $j$, these three measures are defined as
\[
\text{MAE}_j(\mathcal{M}) = \frac{1}{n_{\text{test}}}\sum_{i=1}^{n_\text{test}}\left|\tilde{x}_{ij}-\widehat{\tilde{x}}^{\mathcal{M}}_{ij}\right|,
\]
\[
\text{RMSE}_j(\mathcal{M}) = \sqrt{\frac{1}{n_{\text{test}}}\sum_{i=1}^{n_\text{test}}\left(\tilde{x}_{ij}-\widehat{\tilde{x}}^{\mathcal{M}}_{ij}\right)^2} \,,
\]
and
\[
\text{IS}_j(\mathcal{M}) = \frac{1}{n_{\text{test}}}\sum_{i=1}^{n_\text{test}}\Bigl[\bigl(\widehat{u}^{\mathcal{M}}_{ij}-\widehat{l}^{\mathcal{M}}_{ij} \bigr) + \frac{2}{\alpha}\bigl(\widehat{l}^{\mathcal{M}}_{ij} - \tilde{x}_{ij}\bigr)\mathbb{I}\{\tilde{x}_{ij} < \widehat{l}^{\mathcal{M}}_{ij}\} + \frac{2}{\alpha}\bigl(\tilde{x}_{ij} - \widehat{u}^{\mathcal{M}}_{ij}\bigr)\mathbb{I}\{\tilde{x}_{ij} > \widehat{u}^{\mathcal{M}}_{ij}\}\Bigr],
\]
respectively. 

\begin{figure}[!ht] 
	\centering
	\includegraphics[width=\linewidth]{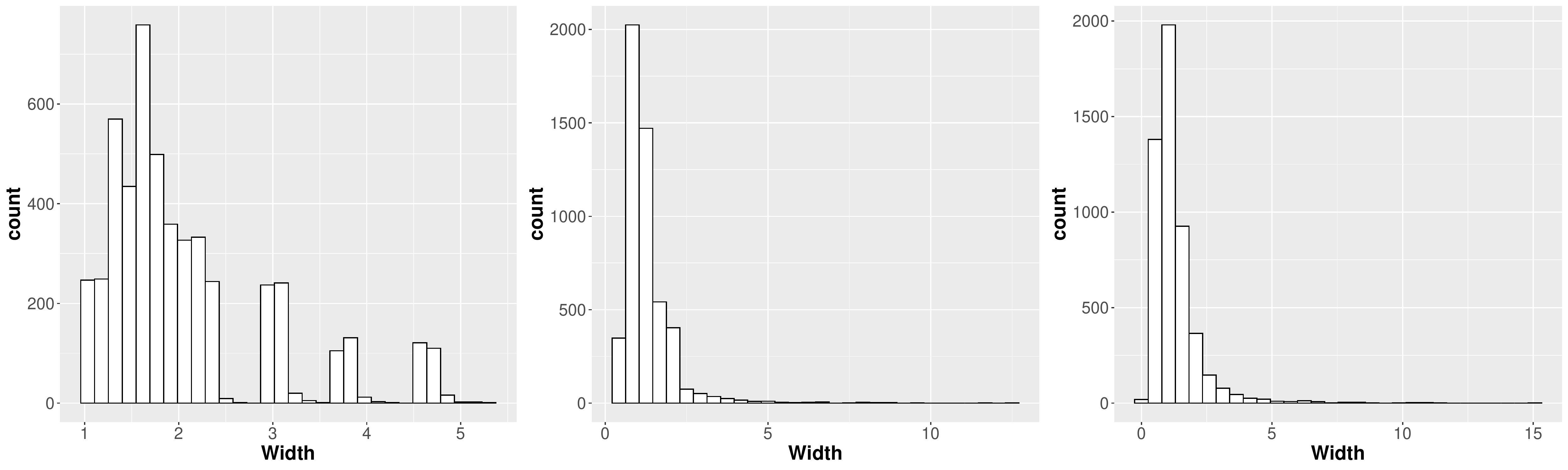}
	\caption{The histograms of the widths of the 80\% prediction intervals generated by linear regression, Gaussian copula cross prediction, and vine copula cross prediction for the credit spread data.} \label{fig: width_3methods}
\end{figure}

Compared with linear regression, cross prediction with Gaussian and vine copulas has smaller MAE, RMSE, and 80\% IS. On one hand, in terms of the interval prediction performance, cross prediction with vine copula has slightly lower 80\% IS than Gaussian copula, while cross prediction with both copula models has much smaller 80\% IS compared with linear regression. The histograms of the widths of the 80\% prediction intervals generated by the three methods are displayed in Figure~\ref{fig: width_3methods}. It can be seen that the widths of the prediction intervals by copula models have a wider range and larger variations compared with linear regression. This implies that heteroscedasticity exists and conditional distributions  are not normally distributed, since univariate marginal distributions of the residual series are not normally distributed. Copula models are thus superior to linear regression in accounting for heteroscedasticity in interval predictions.

\begin{table}[!ht]
	\small
	\centering
	\begin{tabular}{*{10}{l}}
		\toprule
		& \multicolumn{3}{c}{Linear regression} & \multicolumn{3}{c}{Gaussian copula} & \multicolumn{3}{c}{Vine Copula} \\\cmidrule(lr){2-4}\cmidrule(lr){5-7}\cmidrule(lr){8-10}
		Sectors & MAE & RMSE & 80\% IS & MAE & RMSE & 80\% IS & MAE & RMSE & 80\% IS\\\midrule
		City investment&0.384&0.520&2.002&0.333&0.484&1.782&0.308&0.448&1.488\\
		Real estate&0.294&0.454&1.689&0.247&0.410&1.522&0.269&0.445&1.483\\
		Iron&0.629&0.887&3.311&0.656&0.919&3.514&0.668&0.932&3.729\\
		Transportation&0.250&0.326&1.518&0.274&0.358&1.379&0.262&0.367&1.266\\
		Public utilities&0.364&0.483&2.083&0.327&0.447&1.540&0.302&0.399&1.509\\
		Chemical engineering&0.874&1.524&5.997&0.685&1.525&4.335&0.668&1.498&4.535\\
		Construction materials&0.598&0.963&3.574&0.613&0.984&4.217&0.651&1.033&4.223\\
		Construction&0.413&0.641&2.352&0.364&0.587&1.974&0.390&0.643&2.040\\
		Coal&0.702&0.969&3.497&0.674&0.943&3.656&0.678&0.946&3.579\\
		Motor manufacturing&0.703&1.348&5.013&0.658&1.494&4.207&0.648&1.472&3.892\\
		Commerce \& retail&0.162&0.215&1.067&0.156&0.225&0.792&0.152&0.219&0.801\\
		Oil \& gas&0.539&0.830&3.066&0.525&0.822&3.128&0.526&0.821&3.107\\
		Culture \& media&0.521&0.786&3.492&0.430&0.703&2.688&0.433&0.706&2.655\\
		Nonferrous metals&0.625&1.030&3.610&0.606&1.016&3.733&0.630&1.046&3.812\\
		Equipment manufacturing&0.605&1.016&3.392&0.522&0.995&3.182&0.625&1.067&3.640\\
		Electronics&0.623&0.995&3.907&0.572&1.056&3.344&0.549&0.972&3.175\\
		Electricity generation&0.353&0.484&1.979&0.326&0.467&1.764&0.335&0.480&1.715\\
		Electricity supply&0.498&0.738&2.686&0.482&0.675&2.494&0.505&0.760&2.534\\
		National railway&0.768&1.083&4.016&0.735&1.029&3.893&0.751&1.026&3.783\\
		Other comprehensive industries&0.397&0.577&2.214&0.362&0.553&2.123&0.382&0.573&2.157\\\midrule
		Mean&0.515&0.793&3.023&0.477&0.785&2.763&0.486&0.793&2.756\\
		\bottomrule
	\end{tabular}
	\caption{The absolute error (MAE), root mean squared error (RMSE), and interval score (IS) comparing different methods in predicting the ARMA-GARCH filtered residuals after smoothing for different sectors in the credit spread dataset of Chinese corporate bonds.}
	\label{tab: data_performance}
\end{table}

On the other hand, in terms of the point prediction performance, Gaussian copula has slightly smaller MAE and RMSE than vine copula. Since Gaussian copula is a simpler model than vine copula, Gaussian bivariate copulas along with univariate skewed Student's t distribution should be sufficient to predict the median of the residual series if one is interetsed in point prediction. 
The bivariate copulas for fitting a vine copula model on the training set
will be shown in a table in supplementary material.
In tree 1 of the vine, 18 bivariate copulas (15 t, 1 BB1, 1 survival BB1)
have tail dependence, and the other two are BB8 and Frank copulas.
Based on results of \cite{joe2010tail}, the tail dependence in tree 1
indicates there is tail dependence among the most pairs of residuals of the 20 sectors.
In trees 2 and higher, the bivariate copulas are in the
Frank, Gaussian, Gumbel, Clayton, t, BB1 and BB8 families, 
possibly reflected on one variable to get negative dependence or
reflected on two variables to get the opposite tail asymmetry.
For example, the Gumbel copula has only upper tail dependence, so the
survival Gumbel copula has only lower tail dependence.

Although the Gaussian copula model has slightly lower prediction MAE and RMSE, the vine copula provides a better fit on the training data. The overall AIC values of fitting a Gaussian copula and a vine copula on the training set of the credit spread data are 
-4642 and -6004, 
respectively. The overall BIC values of fitting a Gaussian copula and a vine copula on the training set of the credit spread data are 
-3807 and -4870, 
respectively. The AIC and BIC of the vine copula model are both significantly lower than that of the Gaussian copula model, indicating a superior fit on the training data. 
By edges of the vine, we checked that
the AIC values of the vine copula at the first few levels are mostly much smaller when a non-Gaussian copula is fitted. This implies that vine copula model with flexible pair-copulas on edges of the vine fits the training data much better.

\subsection{Risk Transfer} \label{subsec: risk}

To further compare the differences between Gaussian and vine copulas, it is of interest to study how risks transfer among different sectors indicated by the two copula models. This comparison more clearly shows how inferences
are affected when a copula with tail dependence is fitted.

For the evaluation of risk transfer,
we adopt the simulation techniques proposed by \cite{brechmann2013conditional} and details of the algorithmic steps are summarized
in Section \ref{subsec: conditionalsimulation}. 
Specifically, conditioned on an extreme quantile (for example, at 0.95 quantile) of the ARMA-GARCH filtered residuals of the transportation sector, the u-scores of the residuals of all the other sectors 
are simulated based on the conditional distribution derived from the joint copula density. 
We provide a summary of the conditional median u-score values for sectors
within distance 3 of the transportation sector based on the vine graph
(i.e., those sectors paired with the transportation sector
in trees 1, 2 and 3 of the vine).

Using a simulation sample size of 100, u-score vectors are generated
for both Gaussian and vine copulas, conditioned on the 0.95 quantile
of the residual of the transportation sector.  The median residual u-scores of all other sectors can be obtained.  This is repeated 1000
times to get the mean and standard error of the conditional medians;
the results are shown in Table~\ref{tab: risk_sim} for sectors within a distance of 3 of the transportation sector.  Because tail dependence and tail-weighted
dependence are indicated in Table~\ref{tab: dependence_metrics} and almost bivariate copulas in tree 1 have tail dependence, the conditional median u-scores tend to be higher
for the vine copula, showing that risks of a larger magnitude transfer
to other sectors when there is tail dependence.

\begin{table}[!ht]
	\centering
	\begin{tabular}{*{3}{l}}
		\toprule
		Distance-1 & Gaussian copula & Vine copula \\\cmidrule{2-3}
		City investment & 0.860 (0.021) & 0.904 (0.014) \\
		Iron & 0.756 (0.036) & 0.778 (0.035) \\
		Coal & 0.793 (0.031) & 0.819 (0.032) \\
		Commerce \& retail & 0.787 (0.031) & 0.834 (0.030) \\\cmidrule{2-3}
		Mean & 0.799 (0.018) & 0.834 (0.017) \\\midrule
		Distance-2 & Gaussian copula & Vine copula  \\\cmidrule{2-3}
		Real estate & 0.781 (0.033) & 0.815 (0.033) \\
		Public utilities & 0.830 (0.026) & 0.871 (0.021) \\
		Nonferrous metals & 0.773 (0.033) & 0.793 (0.033) \\
		Electricity generation & 0.781 (0.032) & 0.812 (0.032) \\
		Other comprehensive industries & 0.783 (0.031) & 0.829 (0.029)\\\cmidrule{2-3}
		Mean & 0.790 (0.016) & 0.824 (0.016) \\\midrule
		Distance-3 & Gaussian copula & Vine copula  \\\cmidrule{2-3}
		Construction & 0.735 (0.038) & 0.745 (0.040) \\
		Oil \& gas & 0.587 (0.047) & 0.620 (0.046) \\
		Culture \& media & 0.689 (0.043) & 0.663 (0.050) \\
		Equipment manufacturing & 0.715 (0.039) & 0.721 (0.041) \\\cmidrule{2-3}
		Mean & 0.682 (0.023) & 0.687 (0.025) \\
		\bottomrule
	\end{tabular}
	\caption{Simulation from fitted copula models: the means of the conditional medians of the simulated u-scores for the residuals of all sectors within distance 3 of the transportation sector conditioned on 0.95 quantile of the residual for the transportation sector across all repetitions are shown in the table. The numbers in parentheses are the corresponding standard errors.
	The summaries are based on 1000 repetitions with simulation sample size of
	100. 
	}
	\label{tab: risk_sim}
\end{table}

In summary, compared with linear regression, copula models have a larger impact on statistical inference with regard to heteroscedasticity and tail dependence. It is important to take these aspects into account if one would like to generate more informative prediction intervals or study the conditional dependence under circumstances of extreme observations. 

\section{Conclusion} \label{sec: conc}



In conclusion, we have applied vine copula modeling to a credit spread dataset of Chinese corporate bonds. The model is used for conditional inferences
in terms of cross prediction and risk transfer analysis. 
For financial data, the variables often have stronger dependence in the
joint tails than that would be expected with a Gaussian copula model.
The conditional inferences show some types of inferences
are more strongly affected by tail dependence. Risk transfer conditioned on
one variable being extreme shows a big difference between inferences from
a vine copula with tail dependence and a Gaussian copula.
Cross prediction to obtain central prediction interval estimates
for one variable given the others is not a tail inference, and this
explains why the vine and Gaussian copula perform similarly. 
Nevertheless, both copula models provide better prediction intervals than classical 
multiple regression because the variables do not have univariate Gaussian
marginal distributions. The estimated vine structure also provides an interpretable way to visualize the dependence structure among all variables.
In the earlier referenced papers and books where there are comparisons of
vine and Gaussian copulas, there are also tail inferences 
that are affected by tail dependence in the copula model and other inferences that are not affected.

In this paper, the joint distribution of a multivariate dataset is modeled by a complete vine copula. However, in some applications, a copula model with a parsimonious dependence structure is preferred to reduce the number of parameters and avoid overfitting. Future research could be done to modify the proposed cross prediction algorithm when a parsimonious truncated vine copula or a factor copula is fitted instead of a complete vine copula.

\section*{Acknowledgments}
This research has been supported by NSERC Discovery Grant 8698 and Four Year Doctoral Fellowship of the University of British Columbia. We would like to thank the KYZ project at Ping An Asset Management for the assistance on this research.

\bibliographystyle{rfs}
\bibliography{literature}

\appendix
\section*{Appendix}

\renewcommand{\thesubsection}{\Alph{subsection}}
\subsection{Correlation Heatmap of the Filtered Residuals}

\begin{figure}[!ht] 
	\centering
	\includegraphics[width=\linewidth]{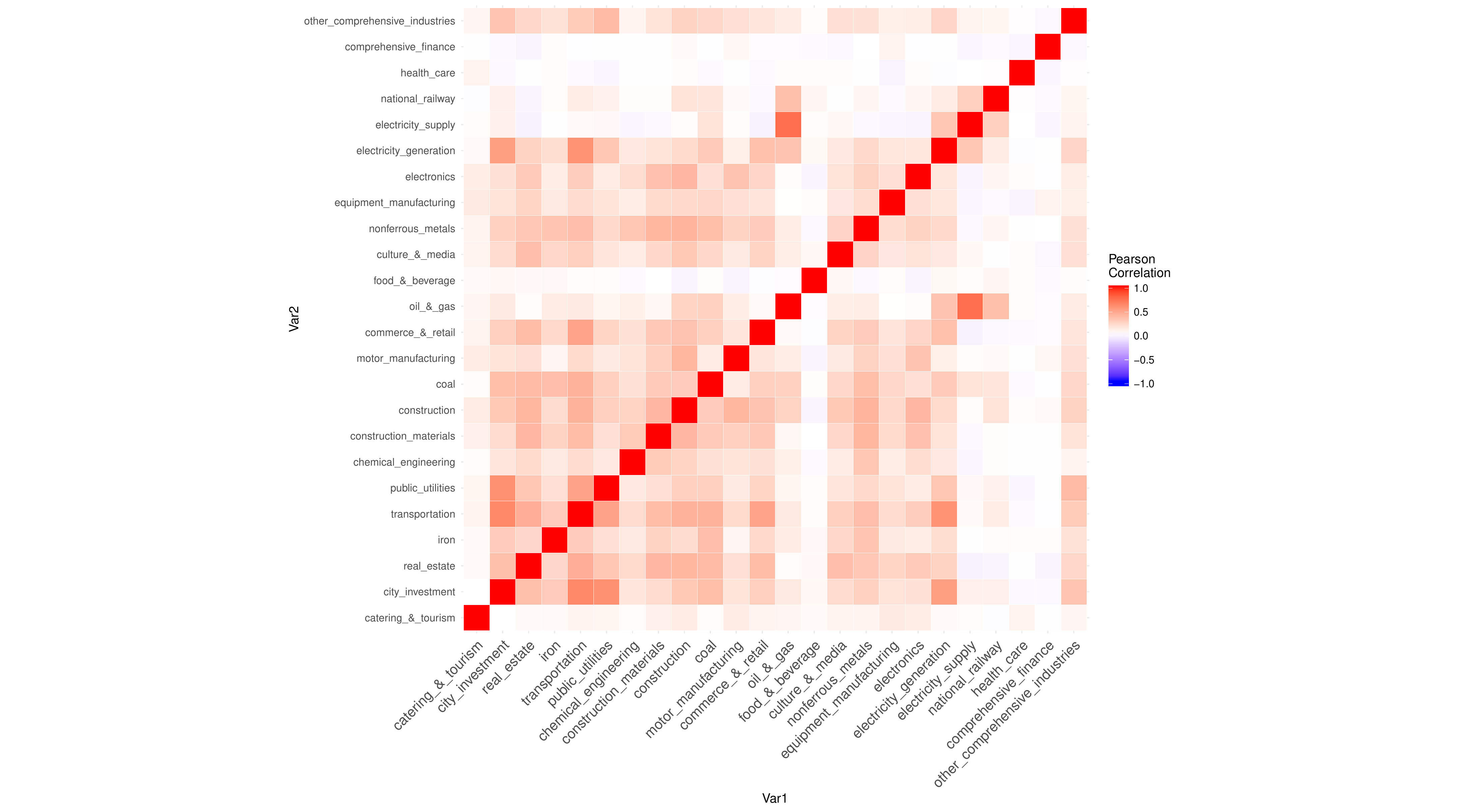}
	\caption{The heatmap of Pearson's correlation coefficient among the obtained residuals after applying the ARMA(1,1)-GARCH(1,1) filter.} \label{fig: heatmap}
\end{figure}

The heatmap of Pearson's correlation coefficient among the unsmoothed ARMA-GARCH filtered residuals of the 24 sectors is shown in Figure~\ref{fig: heatmap}. 

\subsection{Vine Array and Bivariate Copula Families}
The vine array of fitting an R-vine on the training set of the filtered residual series of the credit spread dataset of Chinese corporate bonds is displayed in Table~\ref{tab: vine_array}. 

\begin{table}[!ht]
	\centering
	\begin{tabular}{*{20}{r}}
		\toprule
		1& 1& 4& 4& 4&11&14&15&11&  1& 17& 12& 12&  1& 14&  7& 16&  7&  2&  1\\
		& 4& 1&11& 9& 4&11&14& 4&  4&  1& 17& 18&  4& 11& 15&  7& 16& 11&  5\\
		& &11& 1&11& 9& 4&11&14& 11&  4&  1& 17& 17&  4& 14& 15&  6&  4&  4\\
		& & & 9& 1& 3& 9& 4&15&  9& 11&  4&  1& 12&  2& 11& 14& 15& 14& 17\\
		& & & & 3& 1& 3& 9& 7&  3&  9& 11&  4& 18& 15&  4& 11& 14& 15& 12\\
		& & & & &14& 1& 3& 9& 14&  3&  9& 11& 19&  7&  2&  4& 11&  7& 18\\
		& & & & & &15& 1& 3& 15& 14&  3&  9& 11&  9&  8&  2&  4&  9& 19\\
		& & & & & & & 7& 1&  7& 15& 14&  3&  9&  3&  9&  8&  2&  3& 11\\
		& & & & & & & & 2&  2&  7& 15& 14&  3&  1&  3&  9&  8&  1&  9\\
		& & & & & & & & & 17&  2&  7& 15& 14& 17&  1&  3&  9& 17&  3\\
		& & & & & & & & &  & 12&  2&  7& 15& 12& 17&  1&  3& 12& 14\\
		& & & & & & & & &  &  & 18&  2&  7& 18& 12& 17&  1& 18& 15\\
		& & & & & & & & &  &  &  & 19&  2& 19& 18& 12& 17& 19&  7\\
		& & & & & & & & &  &  &  &  &  5&  5& 19& 18& 12&  5&  2\\
		& & & & & & & & &  &  &  &  &  &  8&  5& 19& 18&  8& 13\\
		& & & & & & & & &  &  &  &  &  &  & 16&  5& 19& 16&  8\\
		& & & & & & & & &  &  &  &  &  &  &  &  6&  5&  6& 16\\
		& & & & & & & & &  &  &  &  &  &  &  &  & 10& 10&  6\\
		& & & & & & & & &  &  &  &  &  &  &  &  &  & 13& 10\\
		& & & & & & & & &  &  &  &  &  &  &  &  &  &  & 20\\
		\bottomrule
	\end{tabular}
	\caption{The vine array of fitting an R-vine on the training set of the filtered residual series of the credit spread dataset of Chinese corporate bonds. The 20 variables are the residual series of the sector-wise mean daily credit spreads of the AAA-rated corporate bonds from the following sectors: (1). city investment, (2). real estate, (3). iron, (4). transportation, (5). public utilities, (6). chemical engineering, (7). construction materials, (8). construction, (9). coal, (10). motor manufacturing, (11). commerce \& retail, (12). oil \& gas, (13). culture \& media, (14). nonferrous metals, (15). equipment manufacturing, (16). electronics, (17). electricity generation, (18). electricity supply, (19). national railway, and (20). other comprehensive industries.}
	\label{tab: vine_array}
\end{table}

The bivariate copulas of fitting a vine copula model on the training set according to the vine array in Table~\ref{tab: vine_array} are shown in Table~\ref{tab: vine_family}. It can be seen that most of the bivariate copula models at the first level are Student's t copula. This indicates that there is tail dependence among the residuals of the 20 sectors. 

\begin{table}[!ht]
	\scriptsize
	\centering
	\begin{tabular}{*{19}{r}}
		\toprule
		t&t&BB1&t&t&t&t&t&t&F&t&BB8&t&t&t&t&BB1.s&t&t\\
		-&t&F&F&F&BB8.s&BB8.s&t&t&BB8.v&F&t&F&BB8&BB8.s&BB8.s&t&G.s&BB8.s\\
		-&-&F&BB8&F&F&F&t&F&C.s&t&t&t&N&F&BB8.s&G&F&BB8\\
		-&-&-&C&F&t&BB8.s&BB8&C&F&C.u&N&N&t&F&F&C.v&F&t\\
		-&-&-&-&F&F&t&BB8.s&BB1.v&BB8.s&F&G&F&C.u&G&BB8.s&BB1.s&t&C.s\\
		-&-&-&-&-&F&F&G&G.v&G.u&F&N&F&F&t&C&N&t&C\\
		-&-&-&-&-&-&F&C.u&t&C.v&F&C.s&t&F&C&G&C.s&F&F\\
		-&-&-&-&-&-&-&N&C.u&C.u&G.u&F&F&F&N&C.s&BB8.s&C.s&BB8.u\\
		-&-&-&-&-&-&-&-&C&F&t&C.u&BB8.u&F&BB1.u&F&t&G.s&t\\
		-&-&-&-&-&-&-&-&-&F&G&G&C.v&G.v&F&F&G.v&C.v&G\\
		-&-&-&-&-&-&-&-&-&-&F&F&F&N&F&N&N&F&BB1\\
		-&-&-&-&-&-&-&-&-&-&-&G.s&t&F&F&G&G&N&t\\
		-&-&-&-&-&-&-&-&-&-&-&-&C.s&F&t&F&G.s&C.v&BB8.s\\
		-&-&-&-&-&-&-&-&-&-&-&-&-&G.s&C.s&C.v&N&N&G.s\\
		-&-&-&-&-&-&-&-&-&-&-&-&-&-&C&G.u&G.s&t&t\\
		-&-&-&-&-&-&-&-&-&-&-&-&-&-&-&F&G&t&F\\
		-&-&-&-&-&-&-&-&-&-&-&-&-&-&-&-&F&G&C\\
		-&-&-&-&-&-&-&-&-&-&-&-&-&-&-&-&-&C.v&G.s\\
		-&-&-&-&-&-&-&-&-&-&-&-&-&-&-&-&-&-&G\\
		\bottomrule
	\end{tabular}
	\caption{The bivariate copulas of fitting an R-vine on the training set of the filtered residual series of the credit spread dataset of Chinese corporate bonds. N, t, C, G, and F in the copula family matrix stand for bivariate Gaussian, Student's t, Clayton, Gumbel, and Frank copulas, respectively. A suffix of `s' represents survival version of the copula family to get the opposite direction of joint tail asymmetry; `u' and `v' represent the copula family with reflection on the first and second variable respectively to get negative dependence.}
	\label{tab: vine_family}
\end{table}

\subsection{Visualization of the Prediction Performance}

\begin{figure}[!ht] 
	\centering
	\includegraphics[width=0.6\linewidth]{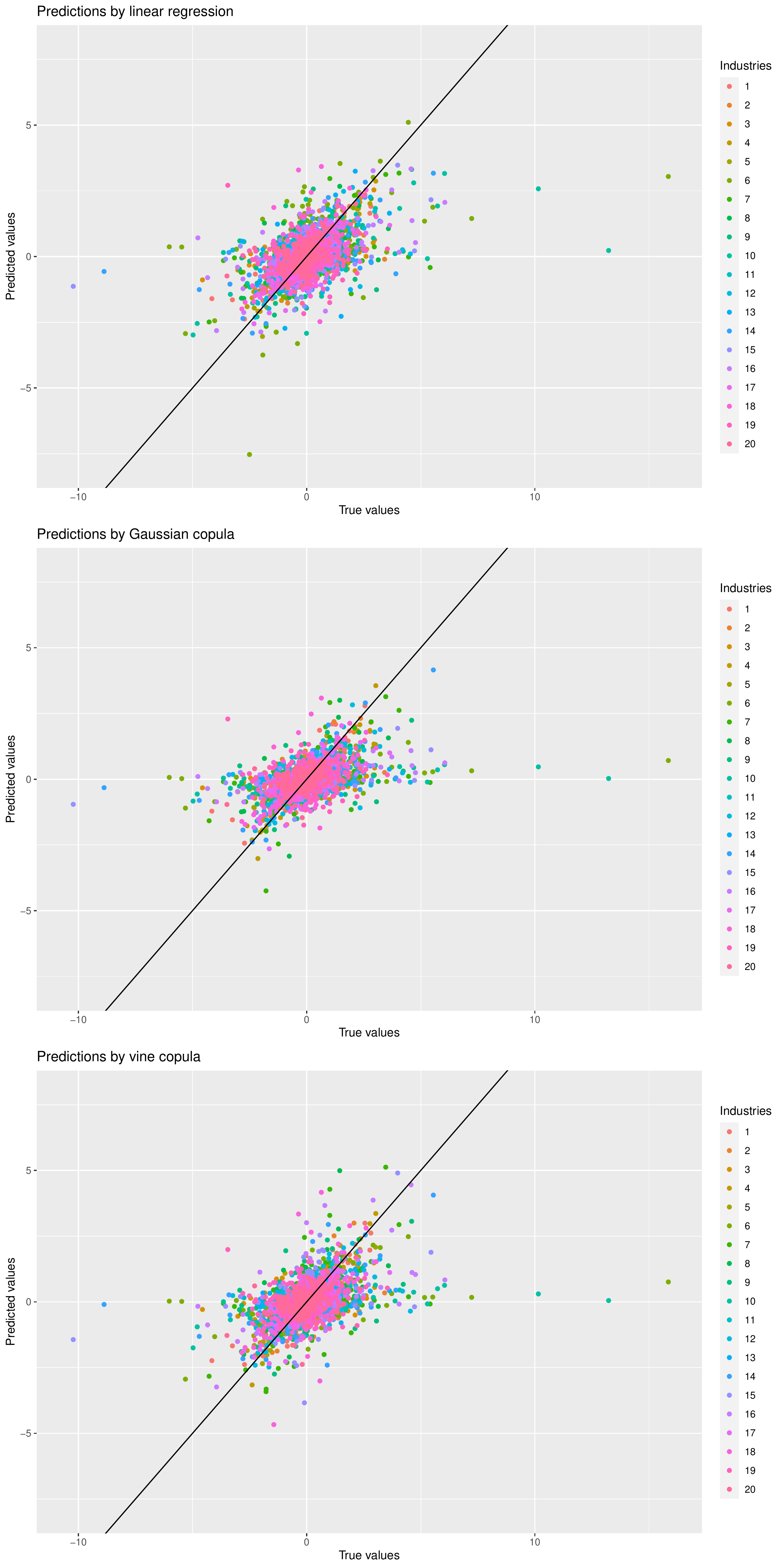}
	\caption{The predicted conditional expectations or medians against the true ARMA-GARCH filtered residuals on the test set by linear regression, Gaussian copula cross prediction, and vine copula cross prediction. The black line is the 45\textdegree diagonal line.} \label{fig: prediction_3methods}
\end{figure}

Figure~\ref{fig: prediction_3methods} visualizes the prediction performance of different methods on the test set. The plots show the predicted conditional expectations or medians against the true values on the test set by linear regression, Gaussian copula cross prediction, and vine copula cross prediction. It can be seen that the point predictions by the three methods are similar, while Gaussian copula and vine copula have slightly lower prediction error.

\subsection{AIC and BIC Values for Copula Models}
The AIC values for each edge when fitting a Gaussian copula on the training set according to the vine array in Table~\ref{tab: vine_array} are shown in Table~\ref{tab: gaussian_AIC}, while the AIC values for each edge when fitting a vine copula on the training set according to the vine array in Table~\ref{tab: vine_array} and copula models in Table~\ref{tab: vine_family} are shown in Table~\ref{tab: vine_AIC}. 

\begin{table}[!ht]
	\tiny
	\centering
	\begin{tabular}{*{19}{r}}
		\toprule
		-331.8&-171.7&-167.5&-112.0&-163.9&-126.4&-159.7&-182.3&-229.0&-125.0&-525.9&-113.8&-328.8&-150.8&-115.8&-86.4&-87.2&-252.2&-251.0\\
		&-22.1&-25.8&-40.3&-52.2&-25.0&-55.0&-67.2&-13.8&-16.0&-12.5&-21.6&-46.2&-20.9&-16.4&-54.0&-17.0&-32.4&-34.9\\
		&&-10.8&-13.3&-35.9&-12.9&-10.1&-22.3&0.6&2.0&-6.9&1.3&-0.2&-18.7&-8.4&-3.0&-2.1&1.9&-6.5\\
		&&&-3.1&-9.3&-10.1&-14.0&-13.9&-4.7&-3.5&0.9&-12.4&-2.4&-5.8&1.7&-25.7&1.9&-6.4&1.1\\
		&&&&1.6&0.4&-3.5&-16.9&-1.3&-24.2&-11.7&-0.2&1.6&1.8&-3.0&-2.5&-2.0&1.8&1.4\\
		&&&&&2.0&-3.1&-4.0&1.2&1.1&-2.1&0.5&1.7&-12.6&-1.5&2.0&0.1&-0.4&2.0\\
		&&&&&&-0.8&1.7&1.5&1.0&2.0&-1.8&2.0&-1.3&-0.4&2.0&2.0&0.2&2.0\\
		&&&&&&&1.3&-1.7&-2.7&-1.1&1.8&2.0&1.9&1.1&0.8&-0.5&-2.7&1.1\\
		&&&&&&&&-0.9&1.9&1.6&2.0&-0.8&1.7&1.3&1.7&-4.1&0.5&1.5\\
		&&&&&&&&&-1.1&2.0&1.9&2.0&1.9&2.0&1.8&0.3&0.6&0.8\\
		&&&&&&&&&&1.8&2.0&-3.1&-72.8&2.0&1.9&1.0&2.0&-10.3\\
		&&&&&&&&&&&2.0&1.2&-25.6&-0.3&1.2&1.7&-2.1&-0.6\\
		&&&&&&&&&&&&-0.8&-13.4&1.6&1.8&1.4&0.9&-2.5\\
		&&&&&&&&&&&&&1.0&2.0&0.3&1.2&-8.9&0.4\\
		&&&&&&&&&&&&&&0.2&2.0&2.0&-0.8&0.7\\
		&&&&&&&&&&&&&&&1.8&1.6&2.0&1.9\\
		&&&&&&&&&&&&&&&&1.0&1.9&2.0\\
		&&&&&&&&&&&&&&&&&-2.1&1.9\\
		&&&&&&&&&&&&&&&&&&0.4\\
		\bottomrule
	\end{tabular}
	\caption{The AIC values for each edge when fitting a Gaussian copula on the training set based on the vine array in Table~\ref{tab: vine_array}.}
	\label{tab: gaussian_AIC}
\end{table}

\begin{table}[!ht]
	\tiny
	\centering
	\begin{tabular}{*{19}{r}}
		\toprule
		-432.9&-242.1&-178.0&-121.7&-224.0&-152.3&-211.0&-253.0&-255.2&-147.6&-58&-123.2&-442.7&-183.4&-222.1&-171.5&-114.5&-318.5&-307.0\\
		&-55.8&-25.1&-55.1&-5&-32.9&-63.5&-65.0&-20.2&-22.3&-16.9&-35.8&-47.2&-22.3&-20.4&-31.7&-40.5&-26.6&-41.9\\
		&&-6.7&-15.9&-49.5&-10.4&-4.0&-23.7&0.4&0.6&-24.0&-1.4&-3.7&-19.1&-5.7&-1.5&-9.1&-0.2&-7.9\\
		&&&-1.8&-6.8&-12.0&-13.2&-7.4&-13.7&-4.1&-0.6&-7.6&-2.6&-11.2&0.7&-24.7&2.0&-9.3&-0.1\\
		&&&&0.1&-1.9&-9.9&-21.9&-1.7&-32.4&-9.8&-0.1&1.3&-0.2&-1.2&-3.9&-3.3&-1.6&-0.6\\
		&&&&&1.1&-4.9&-3.7&1.2&-0.9&-2.8&-0.8&1.8&-14.4&-7.8&1.9&-0.9&-7.5&0.0\\
		&&&&&&-10.8&-1.8&1.3&0.1&1.4&-3.7&-9.2&-3.4&-4.9&0.9&0.4&0.5&1.6\\
		&&&&&&&1.7&-2.4&-2.2&-3.1&1.8&0.6&1.8&0.8&0.4&-9.9&-2.8&-1.4\\
		&&&&&&&&-3.7&1.3&-0.8&1.6&-3.3&1.3&0.5&0.4&-4.9&-0.3&-0.9\\
		&&&&&&&&&-3.7&1.0&0.6&1.0&1.7&1.2&1.4&0.8&0.6&-4.7\\
		&&&&&&&&&&1.7&1.3&-6.2&-69.5&1.6&1.6&1.1&0.7&-12.4\\
		&&&&&&&&&&&1.7&-2.9&-41.2&-0.4&0.1&-3.5&-0.5&1.1\\
		&&&&&&&&&&&&-0.2&-12.4&0.7&1.7&0.3&0.7&-12.8\\
		&&&&&&&&&&&&&-6.2&1.6&0.5&0.6&-4.2&-1.1\\
		&&&&&&&&&&&&&&-7.6&0.0&0.1&-2.2&-1.3\\
		&&&&&&&&&&&&&&&2.0&-1.0&-5.8&-0.2\\
		&&&&&&&&&&&&&&&&-0.1&0.7&1.6\\
		&&&&&&&&&&&&&&&&&-2.4&-1.7\\
		&&&&&&&&&&&&&&&&&&-5.6\\
		\bottomrule
	\end{tabular}
	\caption{The AIC values for each edge when fitting a vine copula on the training set based on the vine array in Table~\ref{tab: vine_array} and the copula models in Table~\ref{tab: vine_family}.}
	\label{tab: vine_AIC}
\end{table}

\end{document}